\newcommand{\typeof}{0}

\documentclass{article}

\usepackage{ifthen}
\newcommand{\condinc}[2]{\ifthenelse{\equal{\typeof}{0}}{#1}{#2}}
\usepackage[utf8]{inputenc}
\usepackage{amsmath,amssymb}
\usepackage{proof}
\usepackage{mathrsfs}
\usepackage{stmaryrd}
\usepackage{xspace}
\usepackage{array}
\usepackage{url}
\condinc{\usepackage{a4wide}}{}

\newcommand{\eg}{\emph{e.g.}\ }
\newcommand{\some}[3]{#1_{#2}, \cdots, #1_{#3}}
\newcommand{\many}[2]{\some{#1}{1}{#2}}

\def\size#1{|#1|}

\newcommand{\NN} {\mathbb{N}}

\newcommand{\RR} {\mathbb{R}}

\newcommand{\union}{\bigcup}

\def\bl#1{\overline{#1}}                          

\def\un#1{\underline{#1}}                         

\def\qi#1{\llparenthesis #1\rrparenthesis}

\newcounter{theo}
\newtheorem{fact}{Fact}
\newtheorem{definition}[theo]{Definition}
\newtheorem{lemma}[theo]{Lemma}

\newtheorem{proposition}[theo]{Proposition}
\newtheorem{corollary}[theo]{Corollary}
\newtheorem{theorem}[theo]{Theorem}
\newenvironment{proof}{\noindent{\bfseries Proof :}}{\hbox{}\hfill$\Box$}

\newcommand{\conone}{\mathbf{c}}

\newcommand{\suc}{\mathbf{s}}
\newcommand{\sucz}{\mathbf{s}_0}
\newcommand{\suco}{\mathbf{s}_1}
\newcommand{\zero}{\mathbf{0}}
\newcommand{\nil}{\mathbf{nil}}

\newcommand{\funone}{\mathtt{f}}
\newcommand{\funtwo}{\mathtt{g}}
\newcommand{\funthree}{\mathtt{h}}
\newcommand{\funfour}{\mathtt{l}}

\newcommand{\append}{\mathtt{append}}

\newcommand{\termone}{t}
\newcommand{\termtwo}{s}

\newcommand{\valone}{v}
\newcommand{\valtwo}{u}
\newcommand{\valthree}{q}

\newcommand{\patone}{p}
\newcommand{\pattwo}{q}

\edef\defone{d}
\newcommand{\eqone}{e}

\edef\varone{x}
\edef\vartwo{y}
\edef\varthree{z}

\newcommand{\progone}{\mbox{\texttt{f}}}
\def\main{\progone}
\newcommand{\prog}{\langle \Variables, \Constructors, \Functions,
         \Equations \rangle}
\newcommand{\longprogone}{\prog}

\newcommand{\Variables}{\mathcal{X}}
\newcommand{\Constructors}{\mathcal{C}}
\newcommand{\Functions}{\mathcal{F}}
\newcommand{\Terms}{\mathcal{T}}

\newcommand{\Patterns}{\mathcal{P}}
\newcommand{\Fundefs}{\mathcal{D}}
\newcommand{\Equations}{\mathcal{E}}

\newcommand{\Freeterms}{\Terms(\Constructors,\Functions,\Variables)}

\newcommand{\Consterms}{\Terms(\Constructors)}

\newcommand{\Blindmap}{\mathcal{B}}

\newcommand{\SubstitutionSet}{\mbox{{$\mathfrak {S}$}}}
\newcommand{\sem}[1]{\llbracket #1 \rrbracket}

\newcommand{\cbv}{\downarrow}
\newcommand{\ccbv}{\Downarrow}

\newcommand{\Treeone}{\mathcal{T}}

\newcommand{\nodeone}{\eta}
\newcommand{\longnodeone}{\langle \funone, \many{\valone}{n} \rangle}
\newcommand{\longnodetwo}{\langle \funtwo, \many{\valtwo}{m} \rangle}

\newcommand{\regleCT}[1]{\stackrel{\mbox{\scriptsize $#1$}}{\leadsto}}

\newcommand{\PPO}{PPO\xspace}
\newcommand{\PPObis}{EPPO\xspace}

\newdimen\proofrulebreadth \proofrulebreadth=.05em
\newdimen\proofdotseparation \proofdotseparation=1.25ex
\newdimen\proofrulebaseline \proofrulebaseline=2ex
\newcount\proofdotnumber \proofdotnumber=3
\let\then\relax
\def\hfi{\hskip0pt plus.0001fil}
\mathchardef\squigto="3A3B
\newif\ifinsideprooftree\insideprooftreefalse
\newif\ifonleftofproofrule\onleftofproofrulefalse
\newif\ifproofdots\proofdotsfalse
\newif\ifdoubleproof\doubleprooffalse
\let\wereinproofbit\relax
\newdimen\shortenproofleft
\newdimen\shortenproofright
\newdimen\proofbelowshift
\newbox\proofabove
\newbox\proofbelow
\newbox\proofrulename
\def\shiftproofbelow{\let\next\relax\afterassignment\setshiftproofbelow\dimen0 }
\def\shiftproofbelowneg{\def\next{\multiply\dimen0 by-1 }%
\afterassignment\setshiftproofbelow\dimen0 }
\def\setshiftproofbelow{\next\proofbelowshift=\dimen0 }
\def\setproofrulebreadth{\proofrulebreadth}

\def\prooftree{
\ifnum  \lastpenalty=1
\then   \unpenalty
\else   \onleftofproofrulefalse
\fi
\ifonleftofproofrule
\else   \ifinsideprooftree
        \then   \hskip.5em plus1fil
        \fi
\fi
\bgroup
\setbox\proofbelow=\hbox{}\setbox\proofrulename=\hbox{}%
\let\justifies\proofover\let\leadsto\proofoverdots\let\Justifies\proofoverdbl
\let\using\proofusing\let\[\prooftree
\ifinsideprooftree\let\]\endprooftree\fi
\proofdotsfalse\doubleprooffalse
\let\thickness\setproofrulebreadth
\let\shiftright\shiftproofbelow \let\shift\shiftproofbelow
\let\shiftleft\shiftproofbelowneg
\let\ifwasinsideprooftree\ifinsideprooftree
\insideprooftreetrue
\setbox\proofabove=\hbox\bgroup$\displaystyle 
\let\wereinproofbit\prooftree
\shortenproofleft=0pt \shortenproofright=0pt \proofbelowshift=0pt
\onleftofproofruletrue\penalty1
}

\def\eproofbit{
\ifx    \wereinproofbit\prooftree
\then   \ifcase \lastpenalty
        \then   \shortenproofright=0pt  
        \or     \unpenalty\hfil         
        \or     \unpenalty\unskip       
        \else   \shortenproofright=0pt  
        \fi
\fi
\global\dimen0=\shortenproofleft
\global\dimen1=\shortenproofright
\global\dimen2=\proofrulebreadth
\global\dimen3=\proofbelowshift
\global\dimen4=\proofdotseparation
\global\count255=\proofdotnumber
$\egroup  
\shortenproofleft=\dimen0
\shortenproofright=\dimen1
\proofrulebreadth=\dimen2
\proofbelowshift=\dimen3
\proofdotseparation=\dimen4
\proofdotnumber=\count255
}

\def\proofover{
\eproofbit 
\setbox\proofbelow=\hbox\bgroup 
\let\wereinproofbit\proofover
$\displaystyle
}%
\def\proofoverdbl{
\eproofbit 
\doubleprooftrue
\setbox\proofbelow=\hbox\bgroup 
\let\wereinproofbit\proofoverdbl
$\displaystyle
}%
\def\proofoverdots{
\eproofbit 
\proofdotstrue
\setbox\proofbelow=\hbox\bgroup 
\let\wereinproofbit\proofoverdots
$\displaystyle
}%
\def\proofusing{
\eproofbit 
\setbox\proofrulename=\hbox\bgroup 
\let\wereinproofbit\proofusing
\kern0.3em$
}

\def\endprooftree{
\eproofbit 
  \dimen5 =0pt
\dimen0=\wd\proofabove \advance\dimen0-\shortenproofleft
\advance\dimen0-\shortenproofright
\dimen1=.5\dimen0 \advance\dimen1-.5\wd\proofbelow
\dimen4=\dimen1
\advance\dimen1\proofbelowshift \advance\dimen4-\proofbelowshift
\ifdim  \dimen1<0pt
\then   \advance\shortenproofleft\dimen1
        \advance\dimen0-\dimen1
        \dimen1=0pt
        \ifdim  \shortenproofleft<0pt
        \then   \setbox\proofabove=\hbox{%
                        \kern-\shortenproofleft\unhbox\proofabove}%
                \shortenproofleft=0pt
        \fi
\fi
\ifdim  \dimen4<0pt
\then   \advance\shortenproofright\dimen4
        \advance\dimen0-\dimen4
        \dimen4=0pt
\fi
\ifdim  \shortenproofright<\wd\proofrulename
\then   \shortenproofright=\wd\proofrulename
\fi
\dimen2=\shortenproofleft \advance\dimen2 by\dimen1
\dimen3=\shortenproofright\advance\dimen3 by\dimen4
\ifproofdots
\then
        \dimen6=\shortenproofleft \advance\dimen6 .5\dimen0
        \setbox1=\vbox to\proofdotseparation{\vss\hbox{$\cdot$}\vss}%
        \setbox0=\hbox{%
                \advance\dimen6-.5\wd1
                \kern\dimen6
                $\vcenter to\proofdotnumber\proofdotseparation
                        {\leaders\box1\vfill}$%
                \unhbox\proofrulename}%
\else   \dimen6=\fontdimen22\the\textfont2 
        \dimen7=\dimen6
        \advance\dimen6by.5\proofrulebreadth
        \advance\dimen7by-.5\proofrulebreadth
        \setbox0=\hbox{%
                \kern\shortenproofleft
                \ifdoubleproof
                \then   \hbox to\dimen0{%
                        $\mathsurround0pt\mathord=\mkern-6mu%
                        \cleaders\hbox{$\mkern-2mu=\mkern-2mu$}\hfill
                        \mkern-6mu\mathord=$}%
                \else   \vrule height\dimen6 depth-\dimen7 width\dimen0
                \fi
                \unhbox\proofrulename}%
        \ht0=\dimen6 \dp0=-\dimen7
\fi
\let\doll\relax
\ifwasinsideprooftree
\then   \let\VBOX\vbox
\else   \ifmmode\else$\let\doll=$\fi
        \let\VBOX\vcenter
\fi
\VBOX   {\baselineskip\proofrulebaseline \lineskip.2ex
        \expandafter\lineskiplimit\ifproofdots0ex\else-0.6ex\fi
        \hbox   spread\dimen5   {\hfi\unhbox\proofabove\hfi}%
        \hbox{\box0}%
        \hbox   {\kern\dimen2 \box\proofbelow}}\doll%
\global\dimen2=\dimen2
\global\dimen3=\dimen3
\egroup 
\ifonleftofproofrule
\then   \shortenproofleft=\dimen2
\fi
\shortenproofright=\dimen3
\onleftofproofrulefalse
\ifinsideprooftree
\then   \hskip.5em plus 1fil \penalty2
\fi
}

\newcommand{\ninfer}[3]
     {\prooftree
          #1 
          \justifies #2
          \using #3
      \endprooftree}

\newcommand{\ptime}{\textsc{Ptime}\xspace}

\newcommand{\precEqFS}{\preceq_{\Functions}}

\newcommand{\precFS}{\prec_{\Functions}}

\newcommand{\egalFS}{\approx_{\Functions}}

\edef\xpoSt{\prec_{rpo}}

\newcommand{\xpoXSt}{\prec^{p}_{rpo}}

\newenvironment{varitemize}
{
\begin{list}{\labelitemii}
{\setlength{\itemsep}{0pt}
 \setlength{\topsep}{0pt}
 \setlength{\parsep}{0pt}
 \setlength{\partopsep}{0pt}
 \setlength{\leftmargin}{15pt}
 \setlength{\rightmargin}{0pt}
 \setlength{\itemindent}{0pt}
 \setlength{\labelsep}{5pt}
 \setlength{\labelwidth}{10pt}
}}
{
 \end{list} 
}

\newcounter{number}

\newenvironment{varenumerate}
{
\begin{list}{\arabic{number}.}
{
  \usecounter{number}
  \setlength{\itemsep}{0pt}
  \setlength{\topsep}{0pt}
  \setlength{\parsep}{0pt}
  \setlength{\partopsep}{0pt}
  \setlength{\leftmargin}{15pt}
  \setlength{\rightmargin}{0pt}
  \setlength{\itemindent}{0pt}
  \setlength{\labelsep}{5pt}
  \setlength{\labelwidth}{15pt}
}}
{
\end{list} 
}

\condinc{}
{\title{On Quasi-Interpretations, Blind Abstractions \\
and Implicit Complexity\thanks{Work
         partially supported by projects CRISS (ACI), NO-CoST (ANR)}}

\author{Patrick Baillot
\and
Ugo Dal Lago\\
Laboratoire d'Informatique de Paris-Nord /CNRS \\
Universit\'e Paris-Nord, France\\
\{pb,dallago,jean-yves.moyen\}@lipn.univ-paris13.fr\\
\and 
Jean-Yves Moyen}}

{\title{On Quasi-Interpretations, Blind Abstractions \\
and Implicit Complexity\thanks{Work
         partially supported by projects CRISS (ACI), NO-CoST (ANR)}}
\author{Patrick Baillot\footnote{LIPN, Universit\'e Paris 13} 
  \and Ugo Dal Lago\footnotemark[\value{footnote}] \and 
  Jean-Yves Moyen\footnotemark[\value{footnote}]}
  \date{}
}

\begin{document}
\maketitle
\begin{abstract}
  Quasi-interpretations are a technique to guarantee complexity bounds
  on first-order functional programs: with termination orderings they
  give in particular a sufficient condition for a program to be
  executable in polynomial time (\cite{MM00}), called here the
  P-criterion.  We study properties of the programs satisfying the
  P-criterion, in order to better understand its intensional expressive
  power.

  Given a program on binary lists, its blind abstraction is the
  non-de\-ter\-mi\-ni\-stic program obtained by replacing lists by their
  lengths (natural numbers). A program is blindly polynomial if its
  blind abstraction terminates in polynomial time. We show that all
  programs satisfying a variant of the P-criterion are in fact blindly
  polynomial.  Then we give two extensions of the P-criterion: one by
  relaxing the termination ordering condition, and the other one (the
  bounded value property) giving a necessary and sufficient condition
  for a program to be polynomial time executable, with memoisation.
\end{abstract}
%
%
%
\section{Introduction}

\textbf{Implicit computational complexity} (ICC) explores machine-free
characterizations of complexity classes, without referring to explicit
resource bounds but instead by seeing these bounds as consequences of
restrictions on program structures.  It has been mainly developed in
the functional programming paradigm, by taking advantage of ideas from
primitive recursion (\cite{Leivant94,BC92}), proof-theory and linear
logic (\cite{Girard98}), rewriting systems or functional programming
(\cite{BMMTCS,Jones99}), type systems (\cite{Hofmann99})\ldots

Usually ICC results include a soundness and a completeness statement:
the first one says that all programs of a given language (or those
satisfying a criterion) admit a certain complexity property, and the
latter one that all \textit{functions} of the corresponding functional
complexity class can be programmed in this language. For instance in the
case of polynomial time complexity the first statement refers to a
polynomial time evaluation of programs, whereas the second, which is of
an \textit{extensional} nature, refers to the class FP of functions
computable in polynomial time. Theorems of this kind have been given for
many systems, like for instance ramified recursion~\cite{BC92}, variants
of linear logic, fragments of functional languages\ldots

\textbf{Expressivity}. This line of work is motivating from a
programming language perspective, because it suggests ways to control
complexity properties of programs, which is a difficult issue because of
its infinitistic nature. However, extensional correspondence with
complexity classes is usually not enough: a programming language (or a
static analysis methodology for it) offering guarantees in terms of
program safety is plausible only if it captures enough interesting and
natural \emph{algorithms}.

This issue has been pointed out by several authors
(\cite{MM00,Hofmann99}) and advances have been made in the direction of
more liberal ICC systems: some examples are type systems for
\textit{non-size-increasing} computation and
quasi-inter\-pre\-ta\-tions.  However it is not always easy to measure
the improvement a new ICC system brings up. Until now this has been
usually illustrated by providing examples.

We think that to compare in a more appropriate way the algorithmic
aspects of ICC systems and in particular to understand their
limitations, specific methods should be developed.  Indeed, what we need
are \emph{sharp results} on the \textit{intensional} expressive power of
existing systems and on the intrinsic limits of implicit complexity as a
way to isolate large (but decidable) classes of programs with bounded
complexity.  For that we aim to establish properties (like necessary
conditions) of the \textit{programs} captured by an ICC system. \par

\textbf{Quasi-interpretations (QI)} can be considered as a static
analysis methodology to infer asymptotic resource bounds for first-order
functional programs. Used with termination orderings they allow to
define various criteria to guarantee either space or time complexity
bounds (\cite{BMMTCS,BMM05,Amadio05,BMP06}). They present several
advantages: the language for which they can be used is simple to use,
and more importantly the class of programs captured by this approach is
large compared to that caught by other ICC systems. Indeed, all
primitive recursive programs from Bellantoni and Cook's function algebra
can be easily proved to have a QI (\cite{MoyenPhd}).  One particularly
interesting criterion, that we will call here the \textit{P-criterion},
says that programs with certain QIs and recursive path orderings can be
evaluated in polynomial time.

In this paper, we focus our attention on QIs, proving a strong necessary
condition for first-order functional programs on lists having a QI.
More precisely, a program transformation called \textit{blind
abstraction} is presented. It consists in collapsing the constructors
for lists to just one, modifying rewriting rules accordingly.  This
produces in general non-confluent programs, the efficiency of which can
be evaluated by considering all possible evaluations of the program.

In general, the blind abstraction of a polytime first-order functional
program on binary lists is not itself polytime: blinding introduces many
paths that are not available in the original program.  However, we show
that under certain assumptions, blinding a program satisfying the
P-criterion with a (uniform) QI always produces a program which is
polytime, independently from non-confluence.

\textbf{Outline.} We first describe the syntax and operational semantics
of programs (Section~\ref{programs}), before termination orderings and
QIs (Section~\ref{sect:ordering}). Then blind abstractions are
introduced (Section~\ref{sect:blindabstractions}) and we give the main
property of the P-criterion w.r.t. blinding in
section~\ref{sect:linearprograms}, with application to safe
recursion. 
Finally we define a generalization
of the previous termination ordering and of QIs (bounded values
property) which also guarantees the P-criterion
(Section~\ref{sect:extendedPPO}).
%

\section{Programs as Term Rewriting Systems}\label{programs}
\subsection{Syntax and Semantics of Programs}\label{syntaxandsemantics}
We consider first-order term rewriting systems (TRS) with disjoint sets
$\Variables$, $\Functions$, $\Constructors$ resp. of variables, function
symbols and constructor symbols.

\begin{definition}[Syntax]
\label{def:syntax}
The sets of terms and the equations are defined by:
$$
\begin{array}{l@{\extracolsep{-5mm}}r@{\extracolsep{0cm}}p{2mm}l}
\texttt{(values)} & \Consterms \ni \valone & ::= &
\conone \ | \ \conone(\many{\valone}{n}) \\
\texttt{(terms)} & \Freeterms \ni \termone & ::= &\conone \ | \ \varone 
\  |  \  \conone(\many{\termone}{n}) \\
& & & | \  \funone(\many{\termone}{n}) \\
\texttt{(patterns)} & \Patterns \ni \patone & ::= &
\conone \ | \ \varone \ | \  \conone(\many{\patone}{n})\  \\
\texttt{(equations)} & \Fundefs \ni \defone & ::= &
\funone(\many{\patone}{n}) \to \termone 
\end{array}
$$ where $\varone \in \Variables$, $\funone \in \Functions$, and
$\conone \in \Constructors$.  We shall use a type writer font for
function symbols and a bold face font for constructors.
\end{definition}

\begin{definition}[Programs]
\label{def:untyped-program}
A program is a tuple $\progone = \prog$ where $\Equations$ is a set of
$\Fundefs$-equations.  Each variable in the right hand side (rhs) of an
equation also appears in the lhs of the same equation. The program has a
main function symbol in $\Functions$, which we shall also call
$\progone$.
\end{definition}


The domain of the computed functions is the constructor algebra
$\Consterms$. A substitution $\sigma$ is a mapping from variables to
terms.  We note \SubstitutionSet\ the set of \textit{constructor
  substitutions}, \emph{i.e.}  substitutions $\sigma$ with range
$\Consterms$.

Our programs are not necessarily deterministic, that is the TRS is not
necessarily confluent. Non-confluent programs correspond to relations
rather than functions. 
\condinc{Notice that we could define a function by, \eg,
choosing the greatest possible result among all the
executions~\cite{DP02}. Since we only consider the complexity
(that is the length of all executions), this is not important here.}{}

Firstly, we consider a call by value semantics which is displayed in
Figure~\ref{fig:cbv}. The meaning of $\termone \cbv \valone$ is that
$\termone$ evaluates to the constructor term $\valone$. A derivation
will be called a \textit{reduction proof}. The program $\progone$
computes a relation $\sem{\main} : \Consterms^n \to \Consterms$ defined
by: for all $\valtwo_i \in \Consterms, \valone \in
\sem{\main}(\many{\valtwo}{n})$ iff there is a derivation for 
$\main(\many{\valtwo}{n}) \cbv \valone$.  
The size $\size{J}$ of a judgement $J = \termone \cbv \valone$ is the
size $\size{\termone}$ of the lhs term.

\begin{figure*}[ht]
\hrule
\begin{gather*}
\ninfer 
{\conone \in \Constructors \quad 
  \termone_i  \cbv \valone_i} 
{  \conone(\many{\termone}{n}) \cbv  
  \conone(\many{\valone}{n})}  
{\text{(Constructor)}} \quad 
\ninfer { \exists j, \termone_j \notin \Consterms \quad \termone_i  \cbv
  \valone_i  \quad \funone(\many{\valone}{n}) \cbv \valone}
{\funone(\many{\termone}{n})  \cbv \valone} {\text{(Split)}}\\
\ninfer { \funone(\many{\patone}{n}) \to r \in \Equations \quad 
\sigma \in \SubstitutionSet\quad 
\patone_i \sigma = \valone_i \quad  r\sigma \cbv \valone} 
{\funone(\many{\valone}{n})  \cbv \valone} {\text{(Function)}}\\
\end{gather*}
\vspace{-12mm}
\caption{Call by value semantics with respect to a program $\prog$} 
\label{fig:cbv}
\hrule
\end{figure*}

\begin{definition}[Active rules]
  \emph{Passive} semantics rules are $\text{Constructor}$ and
  $\text{Split}$. The only \emph{active} rule is
  $\text{Function}$.

  Let $\pi : \termone \cbv \valone$ be a reduction proof. If we have:
  $$\infer{\termtwo = \funtwo(\many{\valtwo}{n}) \cbv \valtwo}{\eqone =
    \funtwo(\many{\pattwo}{n}) \to r \quad \sigma \in \SubstitutionSet
    \quad \pattwo_i\sigma = \valtwo_i \quad r\sigma \cbv \valtwo}$$ then
  we say that term $\termtwo$ (resp. judgment $J = \termtwo \cbv
  \valtwo$) is \emph{active}.  $\eqone$ is the equation \emph{activated}
  by $\termtwo$ (resp. $J$) and $r\sigma$ (resp.  $r\sigma \cbv
  \valtwo$) is the \emph{activation} of $\termtwo$ (resp.  $J$). Other
  judgements (conclusions of (Split) or (Constructor) rules) are called
  \textit{passive}.
\end{definition}

Notice that the set of active terms in a proof $\pi$ is exactly the set
of terms of the form $\funone(\many{\valone}{n})$, where $\valone_i$ are
constructor terms, appearing in $\pi$. Since the program may be non
deterministic, the equation activated by a term $\termtwo$ depends on
the reduction and on the occurrence of $\termtwo$ in $\pi$, and not only
on $\termtwo$.

\begin{lemma}\label{lem:sizeofactivations}
  For each program, there exists a polynomial $p:\NN\rightarrow\NN$ such
  that for any reduction proof $\pi$, active term $\termone$ in it and
  $\termtwo$ the activation of $\termone$, $\size{\termtwo} =
  p(\size{\termone})$. That is, the size of $\termtwo$ is polynomially
  bounded by the size of $\termone$.
\end{lemma}

\condinc{\begin{proof}
  Because there is only a finite set of equations in a program, each of
  them leading to at most a polynomial increase in size.
\end{proof}}{}

A subderivation of a derivation proof is obtained by recursively
taking a judgement and some of its premises. That is, it is a subgraph
of the proof tree that is also a tree (but does not necessarily go down
to the leaves of the proof).

\begin{definition}[Dependences]
  Let $\pi$ be a reduction proof and $J = \termone \cbv \valone$ be a
  passive judgement appearing in it. A \emph{dependence} of $J$ is a
  subderivation of $\pi$ whose root is $J$ and which contains only
  passive judgements.  The maximum dependence $D_J$ is the biggest
  dependence of a judgement, with respect to inclusion.
\end{definition}

The following is an example program that we will use throughout the
paper ($i \in \{0, 1\}$):
$$\begin{array}{l@{\ \to\ }l}
\funone(\suc_0 \suc_i x)  &  \append(\funone(\suc_1 x),
\funone(\suc_1 x))\\
\funone(\suc_1 x) & x \\
\funone(\nil)  & \nil \\
\append(\suc_i x, y) & \suc_i \append(x,y)\\
\append(\nil, y)  &  y \\
\end{array}
$$


Consider the derivation $\pi$ from Figure \ref{Fig:examplederivation}.
Active judgements appearing in $\pi$ are
$\funone(\suc_0\suc_1\nil)\cbv\nil$, $\funone(\suc_1\nil)\cbv\nil$ and
$\append(\nil,\nil)\cbv\nil$. Let
$J=\append(\funone(\suc_1\nil),\funone(\suc_1\nil))\cbv\nil$ be a
passive judgement. Its maximum dependence $D_J$ is the subderivation
containing $J$ itself.
\begin{figure*}[ht]
$$
\infer
{
  \funone(\suc_0\suc_1\nil)\cbv\nil
}
{
  \infer
  {
    \append(\funone(\suc_1\nil),\funone(\suc_1\nil))\cbv\nil
  }
  {
    \infer
    {\funone(\suc_1\nil)\cbv\nil}
    {\nil\cbv\nil}
    &
    \infer
    {\funone(\suc_1\nil)\cbv\nil}
    {\nil\cbv\nil}
    &
    \infer
    {\append(\nil,\nil)\cbv\nil}
    {\nil\cbv\nil}
  }
}
$$
\caption{Example of reduction proof.}\label{Fig:examplederivation}
\end{figure*}
\begin{lemma}
  \label{lem:dependence}
  Let $\pi$ be a reduction proof and $J = \termone \cbv \valone$ be a
  judgement in it.
  \begin{varenumerate}
 \item For each judgement $\termtwo \cbv \valtwo$ in a
    dependence of $J$, $\termtwo$ is a subterm of $\termone$.
  \item The depth of any dependence of $J$ is bounded by the depth of
  $\termone$.
   \item The number of judgements in a dependence of $J$ is bounded by
    the size of $\termone$.
  \end{varenumerate}
\end{lemma}

\condinc{\begin{proof}
  We can either do a quick induction or look at the rules.
  \begin{varenumerate}
 \item Because passive rules only produce proper subterms or active
  terms.
  \item Because each passive rule decreases the depth of the term.
  \item It is the number of subterms of $\termone$.
  \end{varenumerate}
\end{proof}}{}

\begin{proposition}
  \label{cor:bound-dep}
  For all programs, there exists a polynomial $p:\NN\rightarrow\NN$ such
  that for all derivations $\pi$ and all active judgement $J$ in it with
  a passive activation $H$, $\size{D_{H}} \leq p(\size{J})$. That is,
  the size of $D_{H}$ is polynomially bounded by the size of $J$ and the
  polynomial is only dependent on the program.
\end{proposition}
\condinc{\begin{proof}
Obtained by combining the results of Lemmas \ref{lem:sizeofactivations}
and \ref{lem:dependence}.
\end{proof}}{}
\begin{proposition}
  \label{prop:cbv-bound}
  For all programs, there exists a polynomial $p:\NN^2\rightarrow\NN$
  such that for all derivation proof $\pi$, if $A$ is the number of
  active judgements in $\pi$ and $S$ is the maximum size of an active
  judgement then $\size{\pi} \leq p(A, S)$.
\end{proposition}

So, to bound the time of a derivation, it is sufficient to bound the
number and size of active terms, passive terms playing no real role in
it.

Next, we also consider a call by value semantics with memoisation for
confluent programs. The idea is to maintain a cache to avoid recomputing
the same things several times. Each time a function call is performed,
the semantics looks in the cache. If the same call has already been
computed, then the result can be given immediately, else, we need to
compute the corresponding value and store the result in the cache (for
later reuse). The memoisation semantics is displayed on
Figure~\ref{fig:memoisation}. The (Update) rule can only be triggered if
the (Read) rule cannot, that is if there is no $\valone$ such that
$(\funone, \many{\valone}{n}, \valone) \in C$.

Memoisation corresponds to an automation of the algorithmic technique of
dynamic programming.

\begin{figure*}[ht]
\vspace{-5mm}
\hrule
\begin{gather*}
  \ninfer {\conone \in \Constructors \quad \langle C_{i-1}, \termone_i
    \rangle \ccbv \langle C_i, \valone_i \rangle}{\langle C_0,
    \conone(\many{\termone}{n}) \rangle \ccbv \langle C_n,
    \conone(\many{\valone}{n}) \rangle} {\text{(Constructor)}}\\
  \ninfer { \exists j, \termone_j \notin \Consterms \quad  \langle C_{i-1},
    \termone_i \rangle \ccbv \langle C_i,\valone_i \rangle \quad \langle
    C_n, \funone(\many{\valone}{n}) \rangle \ccbv \langle C, \valone
    \rangle} {\langle C_0, \funone(\many{\termone}{n}) \rangle \ccbv
    \langle C, \valone \rangle}{\text{(Split)}}\\
  \ninfer { (\funone,\many{\valone}{n}, \valone) \in C} {\langle C,
    \funone(\many{\valone}{n}) \rangle \ccbv
    \langle C, \valone \rangle} {\text{(Read)}}\\
  \ninfer { \funone(\many{\patone}{n}) \to r \in \Equations \quad
    \sigma\in \SubstitutionSet \quad \patone_i\sigma = \valone_i \quad
    \langle C, r\sigma \rangle \ccbv \langle C',\valone
    \rangle}{\langle C, \funone(\many{\valone}{n}) \rangle \ccbv
    \langle C' \cup (\funone,\many{\valone}{n},\valone), \valone
    \rangle}{\text{(Update)}}
\end{gather*}
\hfill
\vspace{-5mm}
\caption{Call-by-value interpreter with Cache of $\prog$.}
\label{fig:memoisation}
\hrule
\vspace{-5mm}
\end{figure*}

The expression $\langle C,\termone \rangle \ccbv \langle C',\valone
\rangle$ means that the computation of $\termone$ is $\valone$, given
a program $\progone$ and an initial cache $C$.  The final cache $C'$
contains $C$ and each call which has been necessary to complete the
computation.

\begin{definition}[Active rules]
  $\text{Constructor}$ and
  $\text{Split}$ are \emph{passive} semantic rules. 
  $\text{Update}$ is an \emph{active} rule. $\text{Read}$
  is a \emph{semi-active} rule.
\end{definition}

\begin{definition}[Active terms]
  Active terms and judgements, activated equations, activations and
  dependences are defined similarly as for the call by value case.
  Semi-active terms and judgements are similarly defined for semi-active
  rules.
\end{definition}

Lemma~\ref{lem:dependence} still holds, but we also need to bound the
number of semi-active judgements to bound the size of a derivation.
However, since they only lead to leaves in the derivation and the arity
of the derivation tree is bounded (by $k$ the maximum arity of a symbol
in $\Functions \union \Constructors$), there is at most $k$ times more
(Read) rule than the total number of other rules.

\begin{proposition}
  \label{prop:mem-bound} Consider the memoisation semantics of Figure
  \ref{fig:memoisation}. For all programs, there exists a polynomial $P$
  such that for all derivation proof $\pi$, if $A$ is the number of
  active judgements in $\pi$ and $S$ is the maximum size of an active
  judgement then $\size{\pi} \leq P(A, S)$.
\end{proposition}

\condinc{\begin{proof} Proposition~\ref{cor:bound-dep} still allows to
    bound the size of dependences, hence the number of passive
    judgements. Since semi-active judgements form a subset of the set of
    leaves in $\pi$ and since the number of premises of a rule is
    statically bounded (by $k$, the maximum arity of a symbol), the
    number of semi-active judgements is polynomially bounded by the
    number of non-leaves judgements in $\pi$, hence by the number of
    active and passive judgements.
\end{proof}}{}

\begin{lemma}
  \label{lem:semi-active}
  Let $J_1 = \langle C_1, \termone \rangle \ccbv \langle C_1, \valone
  \rangle$ be a semi-active judgement in a proof $\pi : \langle
  \emptyset, \termone \rangle \ccbv \langle C, \valone \rangle$, then
  there exists an active judgement $J_2 = \langle C_2, \termone \rangle
  \ccbv \langle C'_2, \valone \rangle$ in $\pi$.
\end{lemma}

\condinc{\begin{proof} Because the couple can only be in the cache if an
    active judgement put it there.
\end{proof}}{}




\condinc{The naive model where each rules takes unary time to be
  executed is not very realistic with the memoisation semantics. Indeed,
  each (Read) and (Update) rule needs to perform a lookup in the cache
  and this would take time proportional to the size of the cache (and
  the size of elements in it). However, the size of the final cache is
  exactly the number of (Update) rules in the proof (because only
  (Update) modify the cache) and the size of terms in the cache is
  bounded by the size of active terms (only active terms are stored in
  the cache). So Proposition~\ref{prop:mem-bound} yields to a polynomial
  bound on the execution time.  }{The naive model where each rules takes
  unary time to be executed is not very realistic with the memoisation
  semantics, but even if we consider a cost for (Read) and (Update)
  rules, using Prop.~\ref{prop:mem-bound} we still get a polynomial
  bound on execution time (see Appendix \ref{append:costmemoisation}).}

\condinc{ Memoisation cannot be used with non-confluent programs.
  Indeed, the same function call can lead to several different results.
  Several ideas could be used to define a memoisation semantics for
  non-confluent programs, but they all have their problems, hence we
  won't use any of them here and only use memoisation when the program
  is confluent. For sufficient conditions to decide if a program is
  confluent or not, refer, typically, to Huet's work~\cite{Huet80}.
  Here are, nevertheless, several different hints on how to design a
  memoisation semantics for non-confluent programs.
\begin{varitemize}
\item (No lookup): The cache is never used and everything is recomputed
  every time. This is clearly not satisfactory since this is exactly the
  same thing as the cbv semantics.
\item (Cache first): If a function call is in the cache, use it. This is
  clearly not satisfactory because two identical calls will lead to the
  same result even if there was some non confluence involved.
\item (Random lookup): When performing a call, randomly choose between
  using a result in the cache and doing the computation. This is not
  satisfactory because we can choose to always recompute things, hence
  exactly mimicking the cbv semantics and the worst case will be the
  same (no time is gained).
\item (Random lookup with penalties): Same as random lookup, but after
  (re)com\-pu\-ting a function call, check if is was already in the
  cache.  If so, abort (because one should have looked for the result in
  the cache rather than recomputing it). This seems rather satisfactory
  but brings in lots of problems for analysis. In particular, calls
  following different paths but leading to the same result will be
  identified even if they shouldn't.
\end{varitemize}
}{ Memoisation cannot be used with non-confluent programs. Indeed, the
  same function call can lead to several different results. Several
  ideas could be used to define a memoisation semantics for
  non-confluent programs, but they all have their problems (see
  discussion in Appendix \ref{nondetmemoisation}), hence we won't use
  any of them here and only use memoisation when the program is
  confluent. For sufficient conditions to decide if a program is
  confluent or not, refer, typically, to~\cite{Huet80}.  }

\subsection{Call trees, call dags}

\condinc{ Following~\cite{BMMTCS}, we present now call-trees which are a
  tool that we shall use all along. Let $\progone = \prog$ be a program.
  A call-tree gives a static view of an execution and captures all
  function calls.  Hence, we can study dependencies between function
  calls without taking care of the extra details provided by the
  underlying rewriting relation.  }{ Following~\cite{BMMTCS}, we now
  consider call-trees. Given a program $\progone$, a call-tree gives a
  static view of an execution and captures all function calls.  Hence,
  we can study dependencies between function calls without all the
  details of the semantics.  }
\begin{definition}[States]
\label{def:state}
A \emph{state} is a tuple $\longnodeone$ where $\funone$ is a function
symbol of arity $n$ and $\many{\valone}{n}$ are constructor terms.
Assume that $\nodeone_1 = \longnodeone$ and $\nodeone_2 = \longnodetwo$
are two states.  A \emph{transition} is a triplet $\nodeone_1
\regleCT{\eqone} \nodeone_2$ such that:
\begin{varenumerate}
\item $\eqone$ is an equation $\funone(\many{\patone}{n}) \to \termone$
  of $\Equations$,
\item there is a substitution $\sigma$ such that $\patone_i \sigma =
  \valone_i$ for all $1 \leq i \leq n$,
\item there is a subterm $\funtwo(\many{\termtwo}{m})$ of $\termone$
  such that $\termtwo_i \sigma \cbv \valtwo_i$ for all $1 \leq i
  \leq m$.
\end{varenumerate}

\noindent $\regleCT{*}$ is the reflexive transitive closure of
$\cup_{\eqone \in \Equations}\regleCT{\eqone}$.
\end{definition}


\begin{definition}[Call trees]
  Let $\pi : \termone \cbv \valone$ be a reduction proof. Its \emph{call
    trees} is the set of tree $\Theta_{\pi}$ obtained by only keeping
  active terms in $\pi$.
\end{definition}
\condinc{
That is, if $\termone$ is passive:
$$\ninfer{b \in \Functions \union \Constructors \quad \pi_i : \termone_i
  \cbv \valone_i}{b(\many{\termone}{n}) \cbv
  b(\many{\valone}{n})}{\text{(C) or (S)}}$$ Then $\Theta_{\pi} = \union
\Theta_{\pi_i}$.

If $\termone$ is active:
$$\ninfer{\funone(\many{\patone}{n}) \to r \in \Equations \quad \sigma
  \in \SubstitutionSet \quad \patone_i\sigma = \valone_i \quad \rho :
  r\sigma \cbv \valone}{\funone(\many{\valone}{n}) \cbv
  \valone}{\text{(F)}}$$ Then $\Theta_{\pi}$ only contains the tree
whose root is $\longnodeone$ and children are $\Theta_{\rho}$.  }{ That
is, if $\termone$ is passive:

 $\ninfer{b \in \Functions \union
  \Constructors \quad \pi_i : \termone_i \cbv
  \valone_i}{b(\many{\termone}{n}) \cbv b(\many{\valone}{n})}{\text{(C)
    or (S)}},$ then $\Theta_{\pi} = \union \Theta_{\pi_i}$.

If $\termone$ is active:

 \hspace{-4mm} {\small $\ninfer{\funone(\many{\patone}{n}) \to r
  \in \Equations \quad \sigma \in \SubstitutionSet \quad \patone_i\sigma
  = \valone_i \quad \rho : r\sigma \cbv
  \valone}{\funone(\many{\valone}{n}) \cbv \valone}{\text{(F)}}$}, then
$\Theta_{\pi}$ only contains the tree whose root is $\longnodeone$ and
children are $\Theta_{\rho}$.}

When using the semantics with memoisation, the call-dag of a state is
defined similarly to the call-tree, but using a directed acyclic graph
instead of a tree, that is by adding links from (Read)-judgements to the
corresponding (Update)-judgement. Notice that (Read)-judgement are
always leaves of the proof so we do not loose any part of the proof by
doing so.

\begin{definition}[Call dags]
  Let $\pi : \langle C, \termone \rangle \ccbv \langle C', \valone
  \rangle$ be a reduction proof. Its \emph{call trees} is the set of
  trees obtained by keeping only active terms and its \emph{call dag}
  $\Theta_{\pi}$ is obtained by keeping only active terms and replacing
  each semi-active term by a link to the corresponding (via
  Lemma~\ref{lem:semi-active}) active term.
\end{definition}

\condinc{
\begin{fact}[call tree arity]\label{calltreearity}
  Let $f$ be a program. There exists a fixed integer $k$, such that
  given a derivation $\pi$ of a term of the program, and a tree
  $\Treeone$ of $\Theta_{\pi}$, all nodes in $\Treeone$ have at most $k$
  sons.
\end{fact}

\condinc{\begin{proof}
  For each rhs term $r$ of an equation of $f$ consider the number of
  maximal subterms of $r$ with a function as head symbol; let then $k$
  be the maximum of these integers over the (finite) set of equations of
  $f$.
\end{proof}}{}

\begin{lemma}
  Let $\pi$ be a proof, $\Treeone$ be a call-tree (call-dag) in
  $\Theta_{\pi}$ and consider two states $\nodeone = \longnodeone$ and
  $\nodeone' = \longnodetwo$ such that $\nodeone'$ is a child of
  $\nodeone$. Let $\termone = \funone(\many{\valone}{n})$ be the active
  term corresponding to $\nodeone$ and $\termtwo =
  \funtwo(\many{\valtwo}{m})$ be the active term corresponding to
  $\nodeone'$. Let $\eqone$ be the equation activated by $\termone$ in
  $\pi$. Then, $\nodeone \regleCT{\eqone} \nodeone'$.

  Conversely, if $\nodeone \regleCT{\eqone} \nodeone'$ and $\eqone$ is
  activated by $\termone$ then $\nodeone'$ is a child of $\nodeone$ in
  $\Treeone$.
\end{lemma}

\condinc{\begin{proof} Condition $1$ and $2$ correspond to the
    application of the active rule. Condition $3$ correspond to the
    application of several passive rules to get ride of the context and
    evaluate the parameters of $\funtwo$.
\end{proof}}{}

This means that our definition of call trees is equivalent to the one
in~\cite{BMMTCS}. However, we need an alternate definition in order to
deal with non determinism.  }{ Observe that the notion of call-tree
introduced here is equivalent to the one from~\cite{BMMTCS} in the
deterministic case (see Appendix \ref{appendix:calltrees}). }

Call trees and call dags are a tool to easily count the number
of active judgements in a derivation. So, now, in order to apply
Propositions~\ref{prop:cbv-bound} or~\ref{prop:mem-bound} we need to (i)
bound the size (number of nodes) in the call tree or call dag and (ii)
bound the size of the states appearing in the call tree (dag).

\section{Ordering, Quasi-Interpretations}\label{sect:ordering}
\subsection{Termination Orderings}

\begin{definition}[Precedence]
  Let $\progone = \longprogone$ be a program. A \emph{precedence}
  $\precEqFS$ is a partial ordering over $\Functions \union
  \Constructors$. We note $\egalFS$ the associated equivalence relation.
  A precedence is \emph{compatible} with $\progone$ if for each equation
  $\funone(\many{\patone}{n}) \to r$ and each symbol $b$ appearing in
  $r$, $b \precEqFS \funone$. It is \emph{separating} if for each
  $\conone \in \Constructors, \funone \in \Functions$, $\conone \precFS
  \funone$ (that is constructors are the smallest elements of $\precFS$
  while functions are the biggest). It is \emph{fair} is for each
  constructors $\conone, \conone'$ with the same arity, $\conone \egalFS
  \conone'$ and it is \emph{strict} if for each constructors $\conone,
  \conone'$, $\conone$ and $\conone'$ are incomparable.
\end{definition}

\condinc{
Any strict precedence can be canonically extended into a fair
precedence.
}{}
\begin{definition}[Product extension]
  Let $\prec$ be an ordering over a set $S$. Its \emph{product
    extension} is an ordering $\prec^p$ over tuples of elements of $S$
  such that $(\many{m}{k}) \prec^p (\many{n}{k})$ if and only if (i)
  $\forall i, m_i \preceq n_i$ and (ii) $\exists j$ such that $m_j \prec
  n_j$.
\end{definition}

\begin{definition}[\PPO]
\label{def:xpo}
  Given a separating precedence $\precEqFS$, the recursive path ordering
  $\xpoSt$ is defined in Figure~\ref{fig:xpo}.

 If $\precFS$ is strict (resp. fair) and separating, then the ordering is the
  \emph{Product Path Ordering} \PPO (resp.  the
  \emph{extended Product Path Ordering} \PPObis ).
%
\end{definition} 

\condinc{Of course, it is possible to consider other extensions of
  orderings.  Usual choices are the lexicographic extension, thus
  leading to Lexicographic Path Ordering or Multiset extension, leading
  to Multiset path Ordering. It is also possible to add a notion of
  \emph{status} to function~\cite{KL80} indicating with which extension
  the parameters must be compared. This leads to the more general
  Recursive Path Ordering (RPO). However, here we only use the
  (extended) Product Path Ordering so we don't describe others.  }{ This
  is just a particular case of Recursive Path Orderings (RPO), but here
  we will only use the (extended) Product Path Ordering.}
\begin{figure*}[ht]
\hrule
\begin{gather*}
\ninfer{\termtwo = \termone_i \textit{\ or } \termtwo \xpoSt
  \termone_i}{\termtwo \xpoSt \funone(\ldots, \termone_i, \ldots)}{\funone \in
  \Functions \union \Constructors} \qquad
\ninfer{\forall i\ \termtwo_i \xpoSt \funone(\many{\termone}{n})
  \qquad \funtwo \precFS \funone}{g(\many{\termtwo}{m}) \xpoSt
  \funone(\many{\termone}{n})}{\funone, \funtwo \in \Functions \union
  \Constructors}
\\[5mm]
\ninfer{(\many{\termtwo}{n}) \xpoXSt
  (\many{\termone}{n}) \qquad \funone \egalFS \funtwo \qquad \forall
  i\ \termtwo_i \xpoSt
  \funone(\many{\termone}{n})}{\funtwo(\many{\termtwo}{n}) \xpoSt
  \funone(\many{\termone}{n})}{\funone, \funtwo \in \Functions \union
  \Constructors}
\end{gather*}
\vspace{-5mm}
\caption{Definition of $\xpoSt$} 
\label{fig:xpo}
\hrule
\end{figure*}

An equation $l \to r$ is decreasing if we have $r \xpoSt l$. A program
is ordered by $\xpoSt$ if there is a separating precedence on
$\Functions$ such that each equation is decreasing. Recall that $\xpoSt$
guarantees termination (\cite{Der82}).

Notice that in our case, since patterns cannot contain function symbols,
if there is a precedence such that the program is ordered by the
corresponding $\xpoSt$, then there is also a compatible one with the
same condition.

\begin{lemma}
  Let $\progone$ be a program and $\precFS$ be a separating precedence
  compatible with it. Let $\nodeone = \longnodeone$ be a state in a call
  tree (resp. dag) $\Treeone$ and $\nodeone' = \longnodetwo$ be a
  descendant of $\nodeone$ in $\Treeone$. Then $\funtwo \precEqFS
  \funone$
\end{lemma}

\condinc{\begin{proof}
  Because the precedence is compatible with $\progone$.
\end{proof}}{}

\begin{proposition}[Computing by rank]
  \label{prop:comp-rank}
  Let $\progone$ be a program and $\precFS$ be a separating precedence
  compatible with it. Let $\Treeone$ be a call tree (resp. dag) and $\nodeone
  = \longnodeone$ be a node in it. Let $A$ be the maximum number of
  descendants of a node with the same arity:
$$\begin{array}{l}
 A = \max_{\nodeone = \longnodeone \in \Treeone} \#\{\nodeone' =
  \longnodetwo,\\
 \nodeone \text{ is an ancestor of } \nodeone' \text{ and
    } \funtwo \egalFS \funone\}
\end{array}$$
    The size of $\Treeone$ is polynomially bounded by $A$.

\end{proposition}

\condinc{\begin{proof}
  Let $\funone$ be a function symbol. Its rank is $rk(\funone) =
  \max_{\funtwo \precFS \funone} rk(\funtwo) +1$. 

  Let $d$ be the maximum number of function symbols in a rhs of $\progone$ and
  $k$ be the maximum rank. We will prove by induction that there are at most
  $B_i = \sum_{i \leq j \leq k} d^{k-j} \times A^{k-j+1}$ nodes in
  $\Treeone$ at rank $i$.

  The root has rank $k$. Hence, there are at most $A = d^{k-k} A^{k-k+1}
  = B_k$ nodes at rank $k$.

  Suppose that the hypothesis is true for all ranks $j > i$. Each node
  has at most $d$ children. Hence, there are at most $d \sum_{j > i} B_j$
  nodes at rank $i$ whose parent has rank $\neq i$. Each of these nodes
  has at most $A$ descendants at rank $i$, hence there are at most $d
  \times A \times \sum_{j > i} B_j < B_i$ nodes at rank $i$.

  Since $B_i < (k-i+1) \times d^k A^{k+1}$, $\sum B_i$ is polynomially bounded
  in $A$ and so is the size of the call tree (dag).
%
%
%
%
\end{proof}}{}

Thus to bound the number of active rules in a
derivation (hence bound the derivation's size by
Prop.~\ref{prop:cbv-bound} or~\ref{prop:mem-bound}) it suffices to establish the bound rank by rank.

\begin{proposition}
  \label{prop:ppobound}
  Let $\progone$ be a program terminating by \PPO, $\Treeone$ be a call
  dag and $\nodeone = \longnodeone$ be a node in $\Treeone$. The number
  of descendants of $\nodeone$ in $\Treeone$ with the same rank as
  $\nodeone$ is polynomially bounded by $\size{\nodeone}$.
\end{proposition}

\condinc{\begin{proof}
  Because of the termination ordering, if $\nodeone' = \longnodetwo$ is
  a descendant of $\nodeone$ with $\funone \egalFS \funtwo$, then
  $\valtwo_i$ is a subterm of $\valone_i$. There are at most
  $\size{\valone_i}$ such subterms and thus $c \Pi (\size{\valone_i}
  +1)$ possible nodes (where $c$ is the number of functions with the
  same precedence as $\funone$).
\end{proof}}{}

This is  point $(2)$ in the proof of Lemma~$51$
in~\cite{BMMTCS}. 
Notice that it only works on a call dag, because identical
nodes are identified, and not on a call tree.


\subsection{Quasi-interpretations}
We restrict ourselves to additive QIs as defined
in~\cite{BMMTCS}.

\begin{definition}[Assignment]
  An \emph{assignment} of a symbol $b \in \Functions \union
  \Constructors$ whose arity is $n$ is a function $\qi{b} : (\RR)^n \to
  \RR$ such that:
  \begin{description}
  \item[(Subterm)] $\qi{b}(\many{X}{n}) \geq X_i$ for all $1 \leq i
    \leq n$.
\item[(Weak Monotonicity)] $\qi{b}$ is increasing (not strictly) wrt each variable.
  \item[(Additivity)] $\qi{\conone}(\many{X}{n}) \geq \sum_{i=1}^n X_i +a$
    if $\conone \in \Constructors$ (where $a \geq 1$).
  \item[(Polynomial)] $\qi{b}$ is bounded by a polynomial.
  \end{description}
\end{definition}

\condinc{
We extend assignments $\qi{.}$ to terms canonically.  Given a term $t$
with $n$ variables, the assignment $\qi{t}$ is a function $(\RR)^n \to
\RR$ defined by the rules:
\begin{align*}
  \qi{b(\many{\termone}{n})} & =
  \qi{b}(\qi{\termone_1},\cdots, \qi{\termone_n})\\
  \qi{x} & = X
\end{align*}
}{
We extend assignments $\qi{.}$ to terms canonically.  Given a term $t$
with $n$ variables, the assignment $\qi{t}$ is a function $(\RR)^n \to
\RR$ defined by the rules:
$\qi{b(\many{\termone}{n})} =
  \qi{b}(\qi{\termone_1},\cdots, \qi{\termone_n}) \qquad 
  \qi{x}  = X. $
}
Given two functions $f :(\RR)^n \to \RR$ and $g :(\RR)^m \to \RR$ such
that $n \geq m$, we say that $f \geq g$ iff $\forall X_1, \ldots, X_n :
f(X_1,\ldots,X_n) \geq g(X_1,\ldots, X_m)$.

\condinc{
There are some well-known and useful consequences of such definitions.
We have $\qi{s} \geq \qi{t}$ if $t$ is a subterm of $s$.  Then, for
every substitution $\sigma$, $\qi{s} \geq \qi{t}$ implies that
$\qi{s\sigma} \geq \qi{t\sigma}$.
}{
Consequently we have: $\qi{s} \geq \qi{t}$ if $t$ is a subterm of $s$.  Then, for
every substitution $\sigma$, $\qi{s} \geq \qi{t}$ implies that
$\qi{s\sigma} \geq \qi{t\sigma}$.
}

\begin{definition}[Quasi-interpretation]
  A program assignment $\qi{.}$ is an assignment of each program symbol.
  An assignment $\qi{.}$ of a program is a quasi-interpretation (QI) if for
  each equation $l \to r$, \condinc{\begin{align*}
\qi{l} \geq \qi{r}.
\end{align*}
}{
$\qi{l} \geq \qi{r}.$}
\end{definition}
\condinc{In the following, unless explicitly specified, $\qi{.}$ will always
denote a QI and not an assignment.
}{In the following, $\qi{.}$ will always
denote a QI.}
\begin{lemma}\label{lem:QIandsize}
  Let $\valone$ be a constructor term, $\size{\valone} \leq \qi{\valone}
  \leq a \size{\valone}$ for a  constant $a$. 
\end{lemma}
\condinc{\begin{proof}
  By induction. the constant $a$ depends on the constants in the
  QI of constructors.
\end{proof}}{}
\begin{lemma}\label{lem:CTandQI}
Assume $\progone$ has a QI. Let
  $\nodeone = \longnodeone$ and $\nodeone' = \longnodetwo$ be two
  states such that $\nodeone \regleCT{*} \nodeone'$. Then,
  $\qi{\funtwo(\many{\valtwo}{m})} \leq
  \qi{\funone(\many{\valone}{n})}$.
\end{lemma}

\condinc{\begin{proof}
  Because $\funtwo(\many{\valtwo}{m})$ is a subterm of a term obtained
  by reduction from $\funone(\many{\valone}{n})$.
\end{proof}}{}

\begin{corollary}
  \label{cor:qi-bound}
  Let $\progone$ be a program admitting a QI and $\pi
  : \langle C, \termone = \funone(\many{\valone}{n}) \rangle \ccbv
  \langle C', \valone \rangle$ be a derivation. The size of any active
  term in $\pi$ is bounded by $P(\size{\valone_i})$ for a given
  polynomial $P$.
\end{corollary}

\condinc{\begin{proof}
  The size of an active term $s = \funtwo(\many{\valtwo}{m})$ is bounded
  by $m \max \size{\valtwo_i} \leq m \qi{\termtwo}$. By the previous Lemma,
  $\qi{\termtwo} \leq \qi{\termone}$. But by polynomiality of
  QIs, $\qi{\termone} \leq Q(\qi{\valone_i})$. Since
  $\valone_i$ are constructor terms, $\qi{\valone_i} \leq a
  \size{\valone_i}$
\end{proof}}{}

Now, if we combine this bound on the size of active terms together with
the bound on the number of active terms of
Proposition~\ref{prop:ppobound}, we can apply
Proposition~\ref{prop:mem-bound} and conclude that programs terminating
by \PPO and admitting a QI are \ptime
computable. Actually, the converse is also true:

\begin{theorem}[P-criterion, (\cite{BMMTCS})]\label{PcriterionTheorem}
  The set of functions computable by programs that (i) terminate by
  \PPO and (ii) admit a QI, is exactly \ptime.
\end{theorem}

In order to achieve the polynomial bound, it is necessary to use the cbv
semantics with memoisation.

\section{Blind Abstractions of Programs}\label{sect:blindabstractions}

\subsection{Definitions}
Our idea is to associate to a given program $\funone$ an abstract
program $\bl{\funone}$ obtained by forgetting each piece of data and
replacing it by its size as a unary integer.  In this way, even if
$\funone$ is deterministic, the associated $\bl{\funone}$ will in
general not be deterministic.

For that we first define a target language:
\begin{varitemize}
\item variables: $\bl{\Variables}=\Variables$,
\item function symbols: $\bl{\Functions}=\{\bl{\funone}, \funone \in
  \Functions \}$,
\item constructor symbols: $\bl{\Constructors}= \{\suc, \zero \}$ where
  $\suc$ (resp. $\zero$) has arity 1 (resp. 0).
\end{varitemize}

This language defines a set of constructor terms
$\Terms(\bl{\Constructors})$, a set of terms
$\Terms(\bl{\Constructors}, \bl{\Functions}, \bl{\Variables} )$ and
a set of patterns $\bl{\mathcal{P}}$.
 
The \textit{blinding map} is the natural map $\Blindmap: \Terms(
\Constructors, \Functions , \Variables ) \longrightarrow
\Terms(\bl{\Constructors}, \bl{\Functions}, \bl{\Variables} )$ obtained
by replacing constructors of arity 1 with $\suc$, and those of arity 0
by $\zero$. It induces similar maps on constructor terms and patterns.
We will write $\bl{\termone}$ (resp.  $\bl{p}$) for
$\Blindmap(\termone)$ (resp.  $\Blindmap(p)$).

The blinding map extends to \textit{equations} in the expected way: given a
equation $d = p \rightarrow \termone$ of the language $(\Variables,
\Functions, \Constructors)$, we set $\bl{d}=\Blindmap(d)= \bl{p}
\rightarrow \bl{\termone}$.
Finally, given a program $\funone = (\Variables, \Constructors,
\Functions, \Equations)$, its blind image is $\bl{\funone} =
(\bl{\Variables}, \bl{\Constructors}, \bl{\Functions}, \bl{\Equations})$
where the equations are obtained by:
$\bl{\Equations}= \{\bl{d}, d \in \Equations \}$. Observe that 
even if $\funone$ is an orthogonal program, this will not
necessarily be the case of $\bl{\funone}$, because some patterns are
identified by $\Blindmap$.

The denotational semantics of $\bl{\funone}$ can be seen as a relation
over the domain of tally integers.

\subsection{Complexity Definitions}

\begin{definition}[Strongly polynomial]
  We say a non-deterministic program $\funone$ (of arity $n$) is
    \textit{strongly polynomial} if there exists a polynomial
    $p:\NN\rightarrow\NN$ such that for every sequence
    $\many{\valone}{n}$ and any
    $\pi:\funone(\many{\valone}{n})\cbv\valtwo$, it holds that
    $\size{\pi}\leq p(\sum_{i=1}^n|\valone_i|)$.
\end{definition}
Of course similar definitions would also make sense for other complexity
bounds than polynomial.
In the case of a deterministic program, this definition coincides with
that of a polynomial time program (in the model where rewriting steps
are counted as unit step).
\begin{definition}[Blindly polynomial]
  A program $\funone$ is \textit{blindly polynomial} if its blind
  abstraction $\bl{\funone}$ is strongly polynomial.
\end{definition}
 Observe that:
\begin{fact}
  If a program $\funone$ is blindly polynomial, then it is polynomial
  time (with the call-by-value semantics).
\end{fact}
Indeed, it is sufficient to observe that any reduction sequence of
$\funone$ can be mapped by $\Blindmap$ to a reduction sequence of
$\bl{\funone}$.
The converse property is not true. Observe for that 
our running example in Figure~\ref{Figure1}: 
note that $\funone$ terminates in polynomial time
but this is not the case for $\bl{\funone}$. Indeed if we denote
$\un{n}= \underbrace{\suc \dots \suc}_{n} \; \zero$, we have that
$\bl{\funone}(\un{n})$ can be reduced in an exponential number of steps, with a 
$\pi: \bl{\funone}(\un{n}) \cbv \un{2^n}$.
     
\begin{figure*}
\begin{center}
\begin{tabular}{|c|c|}\hline
$\funone$ & $\bl{\funone}$ 
\\\hline
$
\begin{array}{ccc}
      \funone(\suc_0 \suc_i x) & \rightarrow & \append(\funone(\suc_1
      x), \funone(\suc_1 x))\\
      \funone(\suc_1 x) & \rightarrow & x \\
      \funone(\nil) & \rightarrow & \nil \\
      \append(\suc_i x, y) & \rightarrow & \suc_i \append(x,y) \\
      \append(\nil, y) & \rightarrow & y 
\end{array}
$
&
$
\begin{array}{ccc}
    \bl{\funone}(\suc \suc x) & \rightarrow &
    \bl{\append}(\bl{\funone}(\suc x), \bl{\funone}(\suc x))\\
    \bl{\funone}(\suc x) & \rightarrow & x \\
    \bl{\funone}(0) & \rightarrow & 0 \\
    \bl{\append}(\suc x, y) & \rightarrow & \suc \;
    \bl{\append}(x,y)\\
    \bl{\append}(0, y) & \rightarrow & y 
\end{array}
$
\\
\hline
\end{tabular}
\end{center}
\caption{Blind abstraction of our running example}\label{Figure1}
\end{figure*}

Note that the property of being blindly polynomial is indeed a strong
condition, because it means in some sense that the program will
terminate with a polynomial bound for reasons which are indifferent to
the actual content of the input but only depend on its size.

Now we want to discuss the behavior of the blinding map with
respect to criteria on TRS based on recursive path orderings (RPO) and
QIs (\cite{BMMTCS}).
\subsection{Blinding and Recursive Path Orderings}\label{subsect:RPOandabstraction}


\begin{lemma}\label{RPOblindinglemma}
  Let $\progone$ be a program: if $\progone$ terminates by \PPO then
  $\bl{\progone}$ terminates by \PPO.
\end{lemma}

Indeed $\Functions$ and $\bl{\Functions}$ are in one-one correspondence,
and it is easy to observe that:  if $\precFS$ is a
precedence which gives a PPO ordering for $\progone$, then the corresponding
$\prec_{\bl{\Functions}}$ does the same for $\bl{\progone}$.


The converse is not true, see for example Figure~\ref{Figure1} where
the first equation does terminate by \PPO on the blind side but not on the
non-blind side. 
However, we have:
\begin{proposition}
 Let $f$ be a program. The three following statements
are equivalent: (i) $\progone$ terminates by \PPObis, (ii) $\bl{\progone}$ terminates
by \PPObis, (iii)  $\bl{\progone}$ terminates
by \PPO.
\end{proposition}
\condinc{\begin{proof}
 Just observe that on $\Terms(\bl{\Constructors}, \bl{\Functions}, \bl{\Variables} )$, \PPO and \PPObis
coincide, and that $\bl{t} \prec_{\PPObis} \bl{t'}$ implies $t \prec_{\PPObis} t'$.
\end{proof}}{}
\subsection{Blinding and Quasi-interpretations}

Assume the program $\funone$ admits a quasi-interpretation $\qi{.}$.
Then in general this does not imply that $\bl{\funone}$ admits a
quasi-interpretation.
%
%
Indeed one reason why $\qi{.}$ cannot be simply converted into a
quasi-interpretation for $\bl{\funone}$ is because a
quasi-interpretation might in general give different assignments to
several constructors of the same arity, for instance when
$\qi{\suc_0}(X)=X+1$ and $\qi{\suc_1}(X)=X+2$.  Then when considering
$\bl{\funone}$ there is no natural choice for $\qi{\suc}$.

However, in most examples in practice, a restricted class of
quasi-inter\-pre\-ta\-tions is used:

\begin{definition}[Uniform assignments]
  An assignment for $\funone=(\Variables, \Constructors, \Functions,
  \Equations)$ is \textit{uniform} if all constructors of same arity have
  the same assignment:
  for each $\conone, \conone' \in \Constructors$, 
  $\mathit{arity}(\conone)=\mathit{arity}(\conone')$ implies
  $\qi{\conone}=\qi{\conone'}$. A quasi-interpretation of 
  $\funone$ is uniform if it is defined by a
  uniform assignment.
\end{definition}

Now we have:
\begin{proposition}\label{uniformQIblinding}
  The program $\funone$ admits a uniform quasi-interpretation \textit{iff}
  $\bl{\funone}$ admits a quasi-interpretation.
\end{proposition}

\section{Linear Programs and Call-by-Value Evaluation}\label{sect:linearprograms}

Now we want to use the blinding transformation to examine properties
of programs satisfying the P-criterion (Theorem
\ref{PcriterionTheorem}).

\subsection{Definitions and Main Property}

\begin{definition}[Linearity]
  Let $\funone$ be a program terminating by a RPO and $\funtwo$ be a
  function symbol in $\funone$. We say $\funtwo$ is \textit{linear} in
  $\funone$ if, in the rhs term of any equation for $\funtwo$, there is
  at most one occurrence of a function symbol $\funthree$ with same
  precedence as $\funtwo$. The program $\funone$ is linear if all its
  function symbols are linear.
\end{definition}
  
\begin{theorem}\label{claim1}
  Let $\funone$ be a (possibly non deterministic) program which i)
  terminates by PPO, ii) admits a quasi-interpretation, iii) is linear.
  Then $\funone$ is strongly polynomial. 
\end{theorem}

 
Note that the differences with the P-criterion Theorem from
~\cite{BMMTCS} (Theorem \ref{PcriterionTheorem}) are that: the program
here needs not be deterministic, but linearity is assumed for all
function symbols.  As a result the bound holds not only for the
memoisation semantics, but for the plain call-by-value semantics (and
for all execution sequences). Observe that linearity is here a
sufficient condition to avoid the use of memoisation, which is
problematic with non-determinism (see the end of Section
\ref{syntaxandsemantics}). As blinding produces non-deterministic
programs we thus consider blinding of linear programs.

\condinc{\begin{proof}
  The quasi-interpretation provides a bound on the size of active judgements
  via Corollary~\ref{cor:qi-bound}. Linearity of the program ensures that
  the set of descendants of $\nodeone = \longnodeone$ in a call tree
  with the same precedence as $\funone$ is a branch, that is has size
  bounded by its depth. Termination by \PPO ensure that if $\nodeone' =
  \longnodetwo$ is the child of $\nodeone$ with $\funone \egalFS
  \funtwo$ then $\size{\valtwo_i} \leq \size{\valone_i}$ and there is at
  least one $j$ such that $\size{\valtwo_j} < \size{\valone_j}$. So, the
  number of descendants of $\nodeone$ with the same precedence is
  bounded by $\sum \size{\valone_i}$. This bounds the number of active
  judgements by rank.

  So we can now use Propositions~\ref{prop:comp-rank}
  and~\ref{prop:cbv-bound} and conclude that the size (number of rules)
  of any derivation $\pi : \termone \cbv \valone$ is polynomially
  bounded by $\size{\termone}$.
\end{proof}}{}

\begin{proposition}\label{blindPcriterion}
  Let $\funone$ be a (possibly non deterministic) program which i)
  terminates by PPO, ii) admits a \textit{uniform} quasi-interpretation,
  iii) is linear. Then $\funone$ is blindly polynomial. 
\end{proposition}

\condinc{\begin{proof}
  Note that:
  \begin{varitemize}
  \item $\funone$ terminates by PPO, so $\bl{\funone}$ also, by Lemma
    \ref{RPOblindinglemma};
  \item the quasi-interpretation for $\funone$ is uniform, so
    $\bl{\funone}$ admits a quasi-interpretation, by
    Prop. \ref{uniformQIblinding};
  \item $\funone$ is linear, so $\bl{\funone}$ is also linear.
  \end{varitemize}
  So by Theorem~\ref{claim1} we deduce that $\bl{\funone}$ is strongly
  polynomial. Therefore $\funone$ is blindly polynomial.
\end{proof}}{}

\subsection{Bellantoni-Cook Programs}\label{sect:bellantonicook}
\condinc{ Let BC the class of Bellantoni-Cook programs, as defined
  in~\cite{BC92} written in a Term Rewriting System framework as in
  \cite{MoyenPhd}.  Function arguments are separated into either
  \emph{safe} or \emph{normal} arguments, recurrences can only occur
  over normal arguments and their result can only be used in a safe
  position. We use here a semi-colon to distinguish between normal (on
  the left) and safe (on the right) parameters.

\begin{definition}[Bellantoni-Cook programs]
  The class BC is the smallest class of programs containing:
  \begin{varitemize}
  \item (\textbf{Constant}) $\zero$ 
  \item (\textbf{Successors}) $\suc_i(\varone), i \in \{0, 1\}$ 
  \end{varitemize}
  initial functions:
  \begin{varitemize}
  \item (\textbf{Projection}) $\pi_j^{n,m}(\many{\varone}{n};
    \some{\varone}{n+1}{n+m}) \to \varone_j$ 
  \item (\textbf{Predecessor}) $\texttt{p}(; \zero) \to \zero \quad
    p(;\suc_i(\varone)) \to \varone$
  \item (\textbf{Conditional}) $\texttt{C}(;\zero, \varone, \vartwo)
    \to \varone \quad \texttt{C}(;\suc_0, \varone, \vartwo) \to \varone
    \quad \texttt{C}(;\suc_1, \varone, \vartwo) \to \vartwo$ 
  \end{varitemize}
  and is closed by:
  \begin{varitemize}
  \item (\textbf{Safe recursion}) 
    \begin{align*}
      \funone(\zero, \many{\varone}{n} ;
      \many{\vartwo}{m}) \to & \funtwo(\many{\varone}{n} ;
      \many{\vartwo}{m})\\
      \funone(\suc_i(\varthree), \many{\varone}{n} ;
      \many{\vartwo}{m}) \to & \funthree_i(\varthree, \many{\varone}{n} ;
      \many{\vartwo}{m}, \\
      & \funone(\varthree, \many{\varone}{n} ;
      \many{\vartwo}{m})), i \in \{0, 1\}
    \end{align*}
    with $\funtwo, \funthree_i \in$ BC (previously defined) ;
  \item (\textbf{Safe composition}) $$\funone(\many{\varone}{n} ;
    \many{\vartwo}{m}) \to \funtwo(\funthree_1(\many{\varone}{n}),
    \ldots, \funthree_p(\many{\varone}{n}) ;$$ 
    $$\phantom{\funone(\many{\varone}{n} ;) \to}
    \funfour_1(\many{\varone}{n} ; \many{\vartwo}{m}), \ldots,
    \funfour_q(\many{\varone}{n} ; \many{\vartwo}{m}))$$ with $\funtwo,
    \funthree_i, \funfour_j \in $ BC ;
  \end{varitemize}
\end{definition}

It is easy to see that any BC program terminates by \PPO and is linear.

\begin{definition}[Quasi-interpretations for BC-programs]
  A BC-program admits the following quasi-interpretation:
  \begin{varitemize}
  \item $\qi{\zero} = 1$ ;
  \item $\qi{\suc_i}(X) = X+1$ ;
  \item $\qi{\pi}(\many{X}{n+m}) = \max(\many{X}{n+m})$ ;
  \item $\qi{\texttt{p}}(X) =X$ ;
  \item $\qi{\texttt{C}}(X,Y,Z) = \max(X,Y,Z)$ ;
  \end{varitemize} 
  For functions defined by safe recursion of composition,
  $\qi{\funone}(\many{X}{n};\many{Y}{m}) = q_{\funone}(\many{X}{n}) +
  \max(\many{Y}{m})$ with $q_{\funone}$ defined as follows:
  \begin{varitemize}
  \item $q_{\funone}(A,\many{X}{n}) = A(q_{\funthree_0}(A,\many{X}{n})
    + q_{\funthree_1}(A,\many{X}{n})) + q_{\funtwo}(\many{X}{n})$ if
    $\funone$ is defined by safe recursion ;
  \item $q_{\funone}(\many{X}{n}) =
    q_{\funtwo}(q_{\funthree_1}(\many{X}{n}), \ldots,
    q_{\funthree_p}(\many{X}{n})) + \sum_i q_{\funfour_i}(\many{X}{n})$
    if $\funone$ is defined by safe composition.
  \end{varitemize}
\end{definition}
}
{ Let BC the class of Bellantoni-Cook programs, as defined
  in~\cite{BC92}.  Each BC definition can be turned into a linear
  program terminating by PPO, following~\cite{MoyenPhd}. As suggested
  in~\cite{MoyenPhd}, we can build an additive
  quasi-inter\-pre\-ta\-tion for each such program, which happens to be
  uniform. As a consequence, we have: }

\begin{theorem}
  \label{thm:BC-blind}
  If $\funone$ is a program of BC, then $\funone$ is blindly polynomial.
\end{theorem}

\condinc{\begin{proof}
  It is sufficient to observe that if $f$ is a BC program, then it is
  linear and terminates by PPO, and the quasi-interpretation given above
  is uniform. Therefore by Proposition~\ref{blindPcriterion}, $f$ is
  blindly polynomial.
\end{proof}}{}

\condinc{
\section{Semi-lattices of Quasi-Interpretations}\label{sect:semi-lattice}

The study of necessary conditions on programs satisfying the P-criterion
has drawn our attention to uniform quasi-interpretations. This suggests
to consider quasi-interpretations with fixed assignments for constructors
and to examine their properties as a class.

\begin{definition}[Compatible assignments]
  Let $\progone$ be a program and $\qi{.}_1$, $\qi{.}_2$ be two
  assignments for $\progone$. We say that they are \textit{compatible} if
  for any constructor symbol $\conone$ we have:
  $$\qi{\conone}_1=\qi{\conone}_2.$$.
  A family of assignments for $\progone$ is compatible if its elements
  are pairwise compatible.
  We use these same definitions for quasi-interpretations.
\end{definition}

Each choice of assignments for constructors thus defines a
\textit{maximal compatible family} of quasi-interpretations for a
program $\progone$: all quasi-interpretations for $\funone$ which take
these values on $\Constructors$.
 
We consider on assignments the extensional order $\leq$:
$$
\begin{array}{l}
\qi{.}_1 \leq \qi{.}_2 \quad \mbox{ iff }\\
 \quad \forall f \in
\Constructors \cup \Functions, \forall \vec{x} \in ({\RR}^{+})^k,
\qquad \qi{f}_1(\vec{x}) \leq \qi{f}_2(\vec{x}).
\end{array}$$

Given two compatible assignments $\qi{.}_1$, $\qi{.}_2$ we denote by
$\qi{.}_1 \wedge \qi{.}_2$ the assignment $\qi{.}_0$ defined by:
\begin{eqnarray*}
\forall c\in\Constructors.\qi{c}_0&=& \qi{c}_1= \qi{c}_2\\
\forall f\in\Functions.\qi{f}_0&=&\qi{f}_1\wedge\qi{f}_2
\end{eqnarray*} 
where $\alpha\wedge\beta$ denotes 
the greatest lower bound of $\{\alpha,\beta\}$ in
the pointwise order. Then we have:
\begin{proposition}\label{glbof2qi}
  Let $\funone$ be a program and $\qi{.}_1$, $\qi{.}_2$ be two
  quasi-interpretations for it, then $\qi{.}_1 \wedge \qi{.}_2$ is also
  a quasi-interpretation for $\funone$.
\end{proposition}

To establish this Proposition we need intermediary Lemmas. We continue
to denote $\qi{.}_0=\qi{.}_1 \wedge \qi{.}_2$:
\begin{lemma}\label{monotonicitylemma}
  For any $f$ of $\Functions$ we have that $\qi{f}_0$ is monotone and
  satisfies the subterm property.
\end{lemma}
\begin{proof}
  To prove monotonicity, assume $\vec{x}\leq \vec{y}$, for the product
  ordering.  Then, for $i=1$ or $2$: $ \qi{f}_0(\vec{x})=
  \qi{f}_1(\vec{x})\wedge \qi{f}_2(\vec{x}) \leq \qi{f}_i(\vec{x}) \leq
  \qi{f}_i(\vec{y}) $, using monotonicity of $\qi{f}_i$.
  As this is true for $i=1$ and $2$ we thus have: $
  \qi{f}_0(\vec{x})\leq \qi{f}_1(\vec{y})\wedge \qi{f}_2(\vec{y})=
  \qi{f}_0(\vec{y})$.
  It is also easy to check that $\qi{f}_0$ satisfies the subterm
  property.
\end{proof}

\begin{lemma}\label{boundQI0}
  Let $t$ be a term. We have: $\qi{t}_0\leq \qi{t}_i$, for $i=1, 2$.
\end{lemma}
\begin{proof}
  By induction on $t$, using the definition of $\qi{f}_0$ and
  $\qi{c}_0$, and the monotonicity property of Lemma
  \ref{monotonicitylemma}.
\end{proof}

\begin{lemma}\label{conditionrules}
  Let $g(\many{\patone}{n}) \to \termone$ be an equation of the program $f$.
  We have:
  $$ \qi{g}_0\circ (\qi{p_1}_0, \dots, \qi{p_n}_0) \geq \qi{t}_0.$$ 
\end{lemma}
\begin{proof}
  As patterns only contain constructor and variable symbols, and by
  definition of $\qi{.}_0$, if $p$ is a pattern we have: $\qi{p}_0=
  \qi{p}_1= \qi{p}_2$.
  Let $i=1$ or $2$; we have:
  \begin{eqnarray*}
    \qi{g}_i (\qi{p_1}_i(\vec{x}), \dots, \qi{p_n}_i(\vec{x})) &\geq &
    \qi{t}_i(\vec{x}) \quad \mbox{ because $\qi{.}_i$ is a
      quasi-interpretation,}\\
    &\geq & \qi{t}_0(\vec{x}) \quad \mbox{ with Lemma \ref{boundQI0}.} 
  \end{eqnarray*}
  So:
  $$\qi{g}_i (\qi{p_1}_0(\vec{x}), \dots, \qi{p_n}_0(\vec{x})) \geq 
  \qi{t}_0(\vec{x}), \mbox{ as } \qi{p_j}_i= \qi{p_j}_0. $$ Write
  $\vec{y}= (\qi{p_1}_0(\vec{x}), \dots, \qi{p_n}_0(\vec{x}))$.  As
  $\qi{g}_1 (\vec{y}) \geq \qi{t}_0(\vec{x})$ and $\qi{g}_2 (\vec{y})
  \geq \qi{t}_0(\vec{x})$, by definition of $\qi{g}_0$ we get $\qi{g}_0
  (\vec{y}) \geq \qi{t}_0(\vec{x})$, which ends the proof.
\end{proof}

Now we can proceed with the proof of Prop. \ref{glbof2qi}:

\begin{proof} [ Proof of Prop. \ref{glbof2qi}]
  Observe that Lemma \ref{monotonicitylemma} ensures that $\qi{.}_0$
  satisfies the monotonicity and the subterm conditions, and Lemma
  \ref{conditionrules} that it satisfies the condition w.r.t. the equations
  of the program. The conditions for the constructors are also satisfied
  by definition.  Therefore $\qi{.}_0$ is a quasi-interpretation.
\end{proof}

\begin{proposition}
  Let $\funone$ be a program and $\mathcal{Q}$ be a family of compatible
  quasi-interpretations for $\funone$, then $\wedge_{\qi{.} \in
    \mathcal{Q}} \qi{.}$ is a quasi-interpretation for $\funone$.
  Therefore maximal compatible families of quasi-interpretations for
  $\funone$ have an inferior semi-lattice structure for $\leq$.
\end{proposition}
\begin{proof}
  It is sufficient to generalize Lemmas \ref{monotonicitylemma} and
  \ref{conditionrules} to the case of an arbitrary family $\mathcal{Q}$
  and to apply the same argument as for the proof of Prop.
  \ref{glbof2qi}.
\end{proof}}{}

\section{Extending the P-criterion}\label{sect:extendedPPO}
Blind abstraction suggest to consider not only the \PPO ordering from
the P-criterion, but also an extension which is invariant by the
blinding map, the \PPObis ordering (see Subsection
\ref{subsect:RPOandabstraction}). It is thus natural to ask whether
\PPObis enjoys the same property as \PPO . We prove in this section that
with \PPObis we can still bound the size of the call-dag and thus
generalize the P-criterion. Then, we will also consider the
\emph{bounded value property} which is an extension of the notion of QI.
Here, we bound the number of nodes in the call-dag with a given
precedence. Then, Prop.~\ref{prop:comp-rank} bounds the total number of
nodes in the call-dag.

\begin{fact}
  Since we're working over words (unary constructors), patterns are
  either constructors terms (that is, words), or have the form $\patone
  = \suc_1 (\suc_2 \ldots \suc_n(\varone) \ldots)$. In the second case,
  we will write $\patone = \Omega(x)$ with $\Omega = \suc_1 \suc_2
  \ldots \suc_n$.
\end{fact}

The length of a pattern is the length of the corresponding word:
$\size{\patone} = \size{\Omega}$

\begin{proposition}
  \label{prop:reccals}
  In a program terminating by \PPO or \PPObis, the only calls at the
  same precedence that can occur are of the form
  $$\funone(\many{\patone}{n}) \to C[\funtwo_1(\many{\pattwo^1}{m}),
  \ldots, \funtwo_p(\many{\pattwo^p}{l})]$$ where $C[.]$ is some
  context, $\funone \egalFS \funtwo^k$ and $\patone_i, \pattwo^k_j$ are
  patterns. Moreover, each variable appearing in a $\pattwo^k_i$ appears
  in $\patone_i$.
\end{proposition}

\condinc{\begin{proof}
  This is a direct consequence of the termination ordering.
\end{proof}}{}

Since we will only consider individual calls, we will put in the context
all but one of the $\funtwo_k$: $\funone(\many{\patone}{n}) \to
C[\funtwo(\many{\pattwo}{m})]$

\begin{definition}[Production size]
  Let $\progone$ be a program terminating by \PPObis. Let
  $\funone(\many{\patone}{n}) \to C[\funtwo(\many{\pattwo}{m})]$ be a
  call in it where $\funone \egalFS \funtwo$. The \emph{production size}
  of this call is $\max_i \{ \size{\pattwo_i} \}$. The production size
  of an equation is the greatest production size of any call (at the
  same precedence) in it. That is if we have an equation $\eqone =
  \funone(\many{\patone}{n}) \to C[\funtwo_1(\many{\pattwo^1}{m}),
  \ldots, \funtwo_p(\many{\pattwo^p}{l})]$ where $\funone \egalFS
  \funtwo_k$ then its production size is $K_{\eqone} = \max
  \size{\pattwo_j^k}$ The production size of a function symbol
  is the maximum production size of any equation defining a
  function with the same precedence: $K_{\funthree} = \max_{\funtwo
  \egalFS \funthree} \max_{\eqone = \funtwo(\ldots) \to r} K_{\eqone}$
\end{definition}

\begin{definition}[Normality]
  Let $\progone$ be a program. A function symbol $\funthree$ in it is
  \emph{normal} if the patterns in the definitions of functions with the
  same precedence are bigger than its production size:
  $$ 
  \forall \funtwo \egalFS \funthree, \forall
  \funtwo(\many{\pattwo}{m}) \to r \in \Equations, \size{\pattwo_i} \geq
  K_{\funthree}
  $$
  Let $\progone$ be a program. It is \emph{normal} if all function
  symbols in it are normal.
\end{definition}

This means that during recursive calls, every constructor produced at a
given moment will be consumed by the following pattern matching.

\begin{lemma}
  \label{lem:normalize}
  A \PPObis-program can be normalised with an exponential growth
  \emph{in the size of the program}.
\end{lemma}

The exponential is in the difference between the size of the biggest
production and the size of the smallest pattern (with respect to each
precedence).

\condinc{\begin{proof}[Sketch]
  The idea is to extend the small pattern matchings so that their length
  reaches the length of the biggest production. This is illustrated by
  the following example:
  \begin{eqnarray*}
  \funone(\suco(\suco(\suco(\varone))))&\to&\funone(\sucz(\sucz(\varone)))\\
  \funone(\sucz(\varone))&\to&\funone(\varone)
  \end{eqnarray*}
  In this case, the biggest production has size $2$ but the shortest
  pattern matching has only size $1$. We can normalise the program as
  follows: 
  \begin{eqnarray*}
  \funone(\suco(\suco(\suco(\varone))))&\to&\funone(\sucz(\sucz(\varone)))\\
  \funone(\sucz(\sucz(\varone)))&\to&\funone(\sucz(\varone))\\
  \funone(\sucz(\suco(\varone)))&\to&\funone(\suco(\varone))  
  \end{eqnarray*} 
  Even if the process does extend productions as well as patterns, it
  does terminate because only the smallest patterns, hence the smallest
  productions (due to termination ordering) are extended this way.
\end{proof}}{}

Notice that the exponential growth is indeed in the initial size of the
program and does not depend on the size of any input. Since the size of
the call-dag is bounded by the size of the inputs, this does not hamper
the polynomial bound.
The normalization process preserves termination by \PPO and \PPObis,
semantics and does not decrease time complexity. Hence, bounding the
time complexity of the normalized program is sufficient to bound the
time complexity of the initial program.
In the following, we only consider normal programs. 

Let $\progone$ be a program and $\funtwo$ be a function, we will
enumerate all the symbol of same precedence as $\funtwo$ in the rhs of
$\progone$ and label them $\funtwo^1, \ldots, \funtwo^n$. This is simply
an enumeration, not a renaming of the symbols and if a given symbol
appears several times (in several equations or in the same one), it will
be given several labels (one for each occurrence) with this enumeration.
Now, a path in the call-dag staying only at the same precedence as
$\funtwo$ is canonically identified by a word over $\{ \funtwo^1,
\ldots, \funtwo^n \}$.  We write $\nodeone \regleCT{\omega} \nodeone'$
to denote that $\nodeone$ is an ancestor of $\nodeone'$ and $\omega$ is
the path between them.

\begin{lemma}
  Let $\nodeone_1 = \longnodeone$ and $\nodeone_2 = \longnodetwo$ be two
  states such that $\nodeone_1 \regleCT{\eqone} \nodeone_2$ and
  both function symbols have the same precedence. Then $\size{\valone_i}
  \geq \size{\valtwo_i}$ for all $i$ and there exists $j$ such that
  $\size{\valone_j} > \size{\valtwo_j}$.
\end{lemma}

\condinc{\begin{proof}
  This is a consequence of the termination proof by \PPObis.
\end{proof}}{}

\begin{corollary}
  \label{cor:branch}
  Let $\nodeone = \longnodeone$ be a state. Any branch in the call-dag
  starting from $\nodeone$ has at most $n \times (\max
  \size{\valone_i})$ nodes with the same precedence as $\funone$.
\end{corollary}

\begin{lemma}
  \label{lem:commute}
  Suppose that we have labels $\alpha, \beta$ and $\gamma$ and nodes
  such that $\nodeone \regleCT{\alpha} \nodeone_1 \regleCT{\beta}
  \nodeone'_1 \regleCT{\gamma} \nodeone''_1$ and $\nodeone
  \regleCT{\beta} \nodeone_2 \regleCT{\alpha} \nodeone'_2
  \regleCT{\gamma} \nodeone''_2$. Then $\nodeone''_1 = \nodeone''_2$.
\end{lemma}

\condinc{\begin{proof}
  Since labels are unique, the function symbols in $\nodeone''_1$ and
  $\nodeone''_2$ are the same. It is sufficient to show that the $i$th
  components are the same and apply the same argument for the other
  parameters.

  Let $\valthree, \valone, \valone', \valone'', \valtwo, \valtwo',
  \valtwo'' $ be the $i$th parameters of $\nodeone, \nodeone_1,
  \nodeone'_1, \nodeone''_1, \nodeone_2, \nodeone'_2, \nodeone''_2$
  respectively.  Since we're working on words, an equation $\eqone$ has, with
  respect to the $i$th parameter, the form:
  $$\funone(\ldots, \Omega_{\eqone}(x), \ldots) \to C[\funtwo(\ldots,
  \Omega'_{\eqone}(x), \ldots)]$$ and normalization implies that
  $\size{\Omega'_{\eqone}} \leq \size{\Omega_{\eqone}}$ (previous
  lemma).

  So, in our case, we have:
  \begin{eqnarray*}
  \valthree &=& \Omega_{\alpha}(\varone) \to \Omega'_{\alpha}(\varone) =
  \valone = \Omega_{\beta}(\varone') \to \Omega'_{\beta}(\varone')\\
  &=&\valone' = \Omega_{\gamma}(\varone'') \to \Omega'_{\gamma}(\varone'')
  = \valone''\\
  \valthree&=&\Omega_{\beta}(\vartwo) \to \Omega'_{\beta}(\vartwo) =
  \valtwo = \Omega_{\alpha}(\vartwo') \to \Omega'_{\alpha}(\vartwo')\\
  &=&\valtwo' = \Omega_{\gamma}(\vartwo'') \to \Omega'_{\gamma}(\vartwo'')
  = \valtwo''
  \end{eqnarray*}

  Because of normalization, $\size{\Omega_{\beta}} \geq
  \size{\Omega'_{\alpha}}$. Hence $\varone'$ is a suffix of $\varone$,
  itself a suffix of $\valthree$. Similarly, $\varone''$ and
  $\vartwo''$ are suffixes of $\valthree$.

  Since they're both suffixes of the same word, it is sufficient to show
  that they have the same length in order to show that they are
  identical. 
  \begin{eqnarray*}
    \size{\varone} & = & \size{\valthree} - \size{\Omega_{\alpha}}\\
    \size{\varone'} & = & \size{\valone} - \size{\Omega_{\beta}} =
    \size{\valthree} - \size{\Omega_{\alpha}} + \size{\Omega'_{\alpha}}
    - \size{\Omega_{\beta}}\\
    \size{\varone''} & = & \size{\valthree} - \size{\Omega_{\alpha}} +
    \size{\Omega'_{\alpha}} - \size{\Omega_{\beta}} +
    \size{\Omega'_{\beta}} - \size{\Omega_{\gamma}}\\
    \size{\vartwo''} & = & \size{\valthree} - \size{\Omega_{\beta}} +
    \size{\Omega'_{\beta}} - \size{\Omega_{\alpha}} +
    \size{\Omega'_{\alpha}} - \size{\Omega_{\gamma}}
  \end{eqnarray*}

  So $\varone'' = \vartwo''$ and thus $\valone'' = \valtwo''$.
\end{proof}}{}

\begin{corollary}
  \label{cor:commute}
  Let $\omega_1, \omega_2$ be words and $\alpha$ be a label such that:
  $\nodeone \regleCT{\omega_1} \nodeone_1 \regleCT{\alpha} \nodeone'_1$
  and $\nodeone \regleCT{\omega_2} \nodeone_2 \regleCT{\alpha}
  \nodeone'_2$. If $\omega_1$ and $\omega_2$ have the same commutative
  image, then $\nodeone'_1 = \nodeone'_2$.
\end{corollary}

\condinc{\begin{proof}
  This is a generalization of the previous proof, not an induction on
  it. If $\omega_1 = \alpha_1 \ldots \alpha_n$ then the size in the last
  node is:
  $$ \size{\varone'} = \size{\valthree} - \sum \size{\Omega_{\alpha_i}}
  + \sum \size{\Omega'_{\alpha_i}} - \size{\Omega_{\alpha}}$$ which is
  only dependent on the commutative image of $\omega_1$.
\end{proof}}{}

So, when using the semantics with memoisation, the number of nodes (at a
given precedence) in the call-dag is roughly equal to the number of
paths \emph{modulo commutativity}. So any path can be associated with
the vector whose components are the number of occurrences of the
corresponding label in it.

\begin{proposition}
  \label{prop:eppo-cd-bound}
  Let $\Treeone$ be a call-dag and $\nodeone = \longnodeone$ be a node
  in it. Let $I = n \times (\max \size{\valone_j})$ and $M$ be the
  number of functions with the same precedence as $\funone$.  The number
  of descendants of $\nodeone$ in $\Treeone$ with the same precedence as
  $\funone$ is bounded by $(I+1)^M$, that is a polynomial in
  $\size{\nodeone}$.
\end{proposition}

\condinc{\begin{proof}
  Any descendant of $\nodeone$ with the same precedence can be
  identified with a word over $\{\funone^1, \ldots, \funone^M\}$. By
  Corollary~\ref{cor:branch} we know that these words have length at
  most $I$ and by Corollary~\ref{cor:commute} we know that it is
  sufficient to consider these words modulo commutativity.

  Modulo commutativity, words can be identified to vectors with as many
  components as the number of letters in the alphabet and whose sum of
  components is equal to the length of the word.

  Let $D^n_i$ be the number of elements from $\NN^n$ whose sum is
  $i$. This is the number of words of length $i$ over a $n$-ary alphabet
  modulo commutativity.

  Clearly, $D_{i-1}^n \leq D_i^n$ (take all the $n$-uple whose sum is
  $i-1$, add $1$ to the first component, you obtain $D_{i-1}^n$
  different $n$-uple whose sum of components is $i$).

  Now, to count $D_i^n$, proceed as follows: Choose a value $j$ for the
  first component, then you have to find $n-1$ components whose sum is
  $i-j$, there are $D_{i-j}^{n-1}$ such elements.

  $$D_i^n = \sum_{0 \leq j \leq i} D_{i-j}^{n-1} = \sum_{0 \leq j \leq i}
  D_{j}^{n-1} \leq (i+1) \times D_i^{n-1} \leq (i+1)^{n-1} \times D_i^1
  \leq (i+1)^n$$
\end{proof}}{}

\begin{definition}[Bounded Values]
  A program $\funone = (\Variables,\Constructors,\Functions,\Equations)$
  has polynomially bounded values iff for every function symbol $\funtwo
  \in \Functions$, there is a polynomial $p_{\funtwo}: \NN \rightarrow
  \NN$ such that for every state $\eta'$ appearing in a call tree for
  $\eta=(\funtwo, \valone_1, \ldots, \valone_n)$, $|\eta'| \leq
  p_{\funtwo}(\size{\eta})$.
\end{definition}

\begin{theorem}
  The set of functions computed by programs terminating by \PPObis and
   having polynomially bounded values  is exactly \ptime.
\end{theorem}

\condinc{\begin{proof}
    Proposition~\ref{prop:eppo-cd-bound} bounds the size of the call dag
    by rank. Bounded value property bounds the size of nodes in the call
    dag. So we can apply Proposition~\ref{prop:mem-bound} to bound the
    size of any derivation. The converse is obtained from the
    P-criterion.
\end{proof}}{}

\begin{theorem}
  Let $f$ be a deterministic program terminating by \PPObis. Then the
  following two conditions are equivalent:
  \begin{varenumerate}
  \item\label{pnb}
    $f$ has polynomially bounded values;
  \item\label{plt}
    $f$ is polytime in the call-by-value semantics with memoisation.
  \end{varenumerate}
\end{theorem}

\condinc{\begin{proof} The implication $\ref{pnb} \Rightarrow \ref{plt}$
  is proved as follows: Termination by \PPObis provides a polynomial
  bound on the size of the call-dag (by rank) via
  Proposition~\ref{prop:eppo-cd-bound} and the bounded values property
  provides a polynomial bound on the size of nodes in the call
  dag. Hence we can apply Proposition~\ref{prop:mem-bound} and bound the
  size of any derivation $\pi : \langle \emptyset, \termone \rangle
  \ccbv \langle C, \valone \rangle$ by $P(\size{\termone})$ for some
  polynomial $P$.

  For the converse, it is sufficient to see that a state appearing
  in the call dag also appear in the final cache. Since the size of any
  term in the cache is bounded by the size of the proof (because we need
  to perform as many (Constructor) rules as needed to construct the
  term), it is polynomially bounded because the program is polytime.
\end{proof}}{}

\section{Conclusions}

In this paper, blind abstractions of first-order functional
programs have been introduced and exploited in understanding
the intensional expressive power of quasi-interpretations.
In particular, being blindly polytime has been proved to be
a necessary condition in presence of linear programs,
product path-orderings and uniform quasi-interpretations.
This study has lead us to some other interesting results
about the lattice-structure of uniform quasi-interpretations
and the possibility of extending product path-orderings
preserving polytime soundness.

Further work includes investigations on conditions characterizing
the class of programs (or proofs) captured by quasi-interpretation.
In particular, it is still open whether being blindly polytime is
a necessary \emph{and sufficient} conditions for a program
to have a quasi-interpretation (provided some sort of path-ordering
for it exists).

\condinc{\bibliographystyle{plain}}
{\bibliographystyle{latex8}}
\bibliography{blind}

\condinc{}{
\newpage
\appendix
\begin{center} APPENDIX\end{center}

\section{Proofs of Section \ref{programs}}

Proof of Lemma \ref{lem:sizeofactivations}:
\begin{proof}
  Because there is only a finite set of equations in a program, each of
  them leading to at most a polynomial increase in size.
\end{proof}

Proof of Lemma \ref{lem:dependence}:
\begin{proof}
  We can either do a quick induction or look at the rules.
  \begin{varenumerate}
  \item Because each passive rule decreases the depth of the term.
  \item Because passive rules only produce proper subterms or active
  terms.
  \item It is the number of subterms of $\termone$.
  \end{varenumerate}
\end{proof}

Proof of Proposition \ref{cor:bound-dep}:
\begin{proof}
Obtained by combining the results of Lemmas \ref{lem:sizeofactivations}
and \ref{lem:dependence}.
\end{proof}

Proof of Proposition \ref{prop:mem-bound}:

\begin{proof}
  Proposition~\ref{cor:bound-dep} still allows to bound the size of
  dependences, hence the number of passive judgements. Since semi-active
  judgements form a subset of the set of leaves in $\pi$ and since the
  number of premises of a rule is statically bounded (by $k$, the
  maximum arity of a symbol), the number of semi-active judgements is
  polynomially bounded by the number of non-leaves judgements in $\pi$,
  hence by the number of active and passive judgements.
\end{proof}

Proof of Lemma \ref{lem:semi-active}:

\begin{proof}
  Because the couple can only be in the cache if an active judgement put
  it there.
\end{proof}

\section{Semantics}
\subsection{Discussion on cost for memoisation
  semantics}\label{append:costmemoisation}
The naive model where each rules takes unary time to be executed is not
very realistic with the memoisation semantics. Indeed, each (Read) and
(Update) rule needs to perform a lookup in the cache and this would take
time proportional to the size of the cache (and the size of elements in
it). However, the size of the final cache is exactly the number of
(Update) rules in the proof (because only (Update) modify the cache) and
the size of terms in the cache is bounded by the size of active terms
(only active terms are stored in the cache). So
Proposition~\ref{prop:mem-bound} yields to a polynomial bound on the
execution time.

\subsection{Non-determinism and memoisation}\label{nondetmemoisation}
Memoisation cannot be used with non-confluent programs. Indeed, the same
function call can lead to several different results. Several ideas could
be used to define a memoisation semantics for non-confluent programs,
but they all have their problems, hence we won't use any of them here
and only use memoisation when the program is confluent. For sufficient
conditions to decide if a program is confluent or not, refer, typically,
to Huet's work~\cite{Huet80}.

Here are, nevertheless, several different hints on how to design a
memoisation semantics for non-confluent programs.
\begin{varitemize}
\item (No lookup): The cache is never used and everything is recomputed
  every time. This is clearly not satisfactory since this is exactly the
  same thing as the cbv semantics.
\item (Cache first): If a function call is in the cache, use it. This is
  clearly not satisfactory because two identical calls will lead to the
  same result even if there was some non confluence involved.
\item (Random lookup): When performing a call, randomly choose between
  using a result in the cache and doing the computation. This is not
  satisfactory because we can choose to always recompute things, hence
  exactly mimicking the cbv semantics and the worst case will be the
  same (no time is gained).
\item (Random lookup with penalties): Same as random lookup, but after
  (re)com\-pu\-ting a function call, check if is was already in the
  cache.  If so, abort (because one should have looked for the result in
  the cache rather than recomputing it). This seems rather satisfactory
  but brings in lots of problems for analysis. In particular, calls
  following different paths but leading to the same result will be
  identified even if they shouldn't.
\end{varitemize}

\section{Call trees, call dags}\label{appendix:calltrees}
\begin{fact}[call tree arity]\label{calltreearity}
  Let $f$ be a program. There exists a fixed integer $k$, such that
  given a derivation $\pi$ of a term of the program, and a tree
  $\Treeone$ of $\Theta_{\pi}$, all nodes in $\Treeone$ have at most $k$
  sons.
\end{fact}

\begin{proof}
  For each rhs term $r$ of an equation of $f$ consider the number of
  maximal subterms of $r$ with a function as head symbol; let then $k$
  be the maximum of these integers over the (finite) set of equations of
  $f$.
\end{proof}

\begin{lemma}
  Let $\pi$ be a proof, $\Treeone$ be a call-tree (call-dag) in
  $\Theta_{\pi}$ and consider two states $\nodeone = \longnodeone,
  \nodeone' = \longnodetwo$ such that $\nodeone'$ is a child of
  $\nodeone$. Let $\termone = \funone(\many{\valone}{n})$ be the active
  term corresponding to $\nodeone$ and $\termtwo =
  \funtwo(\many{\valtwo}{m})$ be the active term corresponding to
  $\nodeone'$. Let $\eqone$ be the equation activated by $\termone$ in
  $\pi$. Then, $\nodeone \regleCT{\eqone} \nodeone'$.

  Conversely, if $\nodeone \regleCT{\eqone} \nodeone'$ and $\eqone$ is
  activated by $\termone$ then $\nodeone'$ is a child of $\nodeone$
  in $\Treeone$.
\end{lemma}

\begin{proof} Condition $1$ and $2$ correspond to the
    application of the active rule. Condition $3$ correspond to the
    application of several passive rules to get ride of the context and
    evaluate the parameters of $\funtwo$.
\end{proof}

This means that our definition of call trees is equivalent to the one
in~\cite{BMMTCS}. However, we need an alternate definition in order to
deal with non determinism.}

\condinc{}{\section{Proofs of Section \ref{sect:ordering}}

Proof of Proposition \ref{prop:comp-rank}:

\begin{proof}
  Let $\funone$ be a function symbol. Its rank is $rk(\funone) =
  \max_{\funtwo \precFS \funone} rk(\funtwo) +1$. 

  Let $d$ be the maximum number of function symbols in a rhs of $\progone$ and
  $k$ be the maximum rank. We will prove by induction that there are at most
  $B_i = \sum_{i \leq j \leq k} d^{k-j} \times A^{k-j+1}$ nodes in
  $\Treeone$ at rank $i$.

  The root has rank $k$. Hence, there are at most $A = d^{k-k} A^{k-k+1}
  = B_k$ nodes at rank $k$.

  Suppose that the hypothesis is true for all ranks $j > i$. Each node
  has at most $d$ children. Hence, there are at most $d \sum_{j > i} B_j$
  nodes at rank $i$ whose parent has rank $\neq i$. Each of these nodes
  has at most $A$ descendants at rank $i$, hence there are at most $d
  \times A \times \sum_{j > i} B_j < B_i$ nodes at rank $i$.

  Since $B_i < (k-i+1) \times d^k A^{k+1}$, $\sum B_i$ is polynomially bounded
  in $A$ and so is the size of the call tree (dag).
%
%
%
%
\end{proof}

Proof of Proposition \ref{prop:ppobound}:

\begin{proof}
  Because of the termination ordering, if $\nodeone' = \longnodetwo$ is
  a descendant of $\nodeone$ with $\funone \egalFS \funtwo$, then
  $\valtwo_i$ is a subterm of $\valone_i$. There are at most
  $\size{\valone_i}$ such subterms and thus $c \Pi (\size{\valone_i}
  +1)$ possible nodes (where $c$ is the number of functions with the
  same precedence as $\funone$).
\end{proof}

Proof of Lemma \ref{lem:QIandsize}:

\begin{proof}
  By induction. the constant $a$ depends on the constants in the
  QI of constructors.
\end{proof}

Proof of Lemma \ref{lem:CTandQI}:

\begin{proof}
  Because $\funtwo(\many{\valtwo}{m})$ is a subterm of a term obtained
  by reduction from $\funone(\many{\valone}{n})$.
\end{proof}

Proof of Corollary \ref{cor:qi-bound}:

\begin{proof}
  The size of an active term $s = \funtwo(\many{\valtwo}{m})$ is bounded
  by $m \max \size{\valtwo_i} \leq m \qi{\termtwo}$. By the previous Lemma,
  $\qi{\termtwo} \leq \qi{\termone}$. But by polynomiality of
  QIs, $\qi{\termone} \leq Q(\qi{\valone_i})$. Since
  $\valone_i$ are constructor terms, $\qi{\valone_i} \leq a
  \size{\valone_i}$
\end{proof}}

\condinc{}{\section{Proofs of Section \ref{sect:linearprograms}}

Proof of Theorem \ref{claim1}:

\begin{proof}
  The quasi-interpretation provides a bound on the size of active judgements
  via Corollary~\ref{cor:qi-bound}. Linearity of the program ensures that
  the set of descendants of $\nodeone = \longnodeone$ in a call tree
  with the same precedence as $\funone$ is a branch, that is has size
  bounded by its depth. Termination by \PPO ensure that if $\nodeone' =
  \longnodetwo$ is the child of $\nodeone$ with $\funone \egalFS
  \funtwo$ then $\size{\valtwo_i} \leq \size{\valone_i}$ and there is at
  least one $j$ such that $\size{\valtwo_j} < \size{\valone_j}$. So, the
  number of descendants of $\nodeone$ with the same precedence is
  bounded by $\sum \size{\valone_i}$. This bounds the number of active
  judgements by rank.

  So we can now use Propositions~\ref{prop:comp-rank}
  and~\ref{prop:cbv-bound} and conclude that the size (number of rules)
  of any derivation $\pi : \termone \cbv \valone$ is polynomially
  bounded by $\size{\termone}$.
\end{proof}

Proof of Proposition \ref{blindPcriterion}:

\begin{proof}
  Note that:
  \begin{varitemize}
  \item $\funone$ terminates by PPO, so $\bl{\funone}$ also, by Lemma
    \ref{RPOblindinglemma};
  \item the quasi-interpretation for $\funone$ is uniform, so
    $\bl{\funone}$ admits a quasi-interpretation, by
    Prop. \ref{uniformQIblinding};
  \item $\funone$ is linear, so $\bl{\funone}$ is also linear.
  \end{varitemize}
  So by Theorem~\ref{claim1} we deduce that $\bl{\funone}$ is strongly
  polynomial. Therefore $\funone$ is blindly polynomial.
\end{proof}}

\condinc{}{
\section{Definitions and Proofs from Section~\ref{sect:bellantonicook}}
Let BC the class of Bellantoni-Cook programs, as defined in~\cite{BC92}
written in a Term Rewriting System framework as in \cite{MoyenPhd}.
Function arguments are separated into either \emph{safe} or
\emph{normal} arguments, recurrences can only occur over normal
arguments and their result can only be used in a safe position. We use
here a semi-colon to distinguish between normal (on the left) and safe
(on the right) parameters.

\begin{definition}[Bellantoni-Cook programs]
The class BC is the smallest class of programs containing:
\begin{varitemize}
\item (\textbf{Constant}) $\zero$
\item (\textbf{Successors}) $\suc_i(\varone), i \in \{0, 1\}$
\end{varitemize}
initial functions:
\begin{varitemize}
\item (\textbf{Projection}) $\pi_j^{n,m}(\many{\varone}{n};
  \some{\varone}{n+1}{n+m}) \to \varone_j$
\item (\textbf{Predecessor}) $\texttt{p}(; \zero) \to \zero \quad
  p(;\suc_i(\varone)) \to \varone$
\item (\textbf{Conditional}) $\texttt{C}(;\zero, \varone, \vartwo)
  \to \varone \quad \texttt{C}(;\suc_0, \varone, \vartwo) \to \varone
  \quad \texttt{C}(;\suc_1, \varone, \vartwo) \to \vartwo$
\end{varitemize}
and is closed by:
\begin{varitemize}
\item (\textbf{Safe recursion}) 
  $$
  \begin{array}{l}
    \funone(\zero, \many{\varone}{n};\many{\vartwo}{m}) \to\\
       \;\;\;\funtwo(\many{\varone}{n} ;\many{\vartwo}{m})\\
    \funone(\suc_i(\varthree), \many{\varone}{n}; \many{\vartwo}{m})\to\\
       \;\;\;\funthree_i(\varthree,\many{\varone}{n};\many{\vartwo}{m},\\
       \;\;\;\;\;\;\;\;\funone(\varthree, \many{\varone}{n};\many{\vartwo}{m})), i \in \{0, 1\}
  \end{array}
  $$
  with $\funtwo, \funthree_i \in$ BC (previously defined) ;
\item (\textbf{Safe composition}) 
  $$
  \begin{array}{l}
  \funone(\many{\varone}{n};\many{\vartwo}{m}) \to\\
  \;\;\;\funtwo(\funthree_1(\many{\varone}{n}),\ldots, \funthree_p(\many{\varone}{n}) ;\\ 
  \;\;\;\;\;\;\funfour_1(\many{\varone}{n} ; \many{\vartwo}{m})\\
  \;\;\;\;\;\;,\ldots,\\
  \;\;\;\;\;\;\funfour_q(\many{\varone}{n} ; \many{\vartwo}{m}))
  \end{array}
  $$ 
  with $\funtwo,\funthree_i, \funfour_j \in $ BC ;
\end{varitemize}
\end{definition}

It is easy to see that any BC program terminates by \PPO and is linear.

\begin{definition}[Quasi-interpretations for BC-programs]
A BC-program admits the following quasi-interpretation:
\begin{varitemize}
\item $\qi{\zero} = 1$ ;
\item $\qi{\suc_i}(X) = X+1$ ;
\item $\qi{\pi}(\many{X}{n+m}) = \max(\many{X}{n+m})$ ;
\item $\qi{\texttt{p}}(X) =X$ ;
\item $\qi{\texttt{C}}(X,Y,Z) = \max(X,Y,Z)$ ;
\end{varitemize} 
For functions defined by safe recursion of composition,
$\qi{\funone}(\many{X}{n};\many{Y}{m}) = q_{\funone}(\many{X}{n}) +
\max(\many{Y}{m})$ with $q_{\funone}$ defined as follows:
\begin{varitemize}
\item $q_{\funone}(A,\many{X}{n}) = A(q_{\funthree_0}(A,\many{X}{n})
  + q_{\funthree_1}(A,\many{X}{n})) + q_{\funtwo}(\many{X}{n})$ if
  $\funone$ is defined by safe recursion ;
\item $q_{\funone}(\many{X}{n}) =
  q_{\funtwo}(q_{\funthree_1}(\many{X}{n}), \ldots,
  q_{\funthree_p}(\many{X}{n})) + \sum_i q_{\funfour_i}(\many{X}{n})$
  if $\funone$ is defined by safe composition.
\end{varitemize}
\end{definition}

\newcounter{tmp}
\setcounter{tmp}{\value{theo}}
\setcounter{theo}{36}

\begin{theorem}
  If $\funone$ is a program of BC, then $\funone$ is blindly polynomial.
\end{theorem}
\begin{proof}
  It is sufficient to observe that if $f$ is a BC program, then it is
  linear and terminates by PPO, and the quasi-interpretation given above
  is uniform. Therefore by Proposition~\ref{blindPcriterion}, $f$ is
  blindly polynomial.
\end{proof}
\setcounter{theo}{\value{tmp}}
}

\condinc{}{
\section{Proofs of Section~\ref{sect:extendedPPO}}
Sketch of proof for Lemma~\ref{lem:normalize}:
\begin{proof}
  The idea is to extend the small pattern matchings so that their length
  reaches the length of the biggest production. This is illustrated by
  the following example:
  \begin{eqnarray*}
  \funone(\suco(\suco(\suco(\varone))))&\to&\funone(\sucz(\sucz(\varone)))\\
  \funone(\sucz(\varone))&\to&\funone(\varone)
  \end{eqnarray*}
  In this case, the biggest production has size $2$ but the shortest
  pattern matching has only size $1$. We can normalise the program as
  follows: 
  \begin{eqnarray*}
  \funone(\suco(\suco(\suco(\varone))))&\to&\funone(\sucz(\sucz(\varone)))\\
  \funone(\sucz(\sucz(\varone)))&\to&\funone(\sucz(\varone))\\
  \funone(\sucz(\suco(\varone)))&\to&\funone(\suco(\varone))  
  \end{eqnarray*} 
  Even if the process does extend productions as well as patterns, it
  does terminate because only the smallest patterns, hence the smallest
  productions (due to termination ordering) are extended this way.
\end{proof}

Proof of Lemma~\ref{lem:commute}:
\begin{proof}
  Since labels are unique, the function symbols in $\nodeone''_1$ and
  $\nodeone''_2$ are the same. It is sufficient to show that the $i$th
  components are the same and apply the same argument for the other
  parameters.

  Let $\valthree, \valone, \valone', \valone'', \valtwo, \valtwo',
  \valtwo'' $ be the $i$th parameters of $\nodeone, \nodeone_1,
  \nodeone'_1, \nodeone''_1, \nodeone_2, \nodeone'_2, \nodeone''_2$
  respectively.  Since we're working on words, an equation $\eqone$ has, with
  respect to the $i$th parameter, the form:
  $$\funone(\ldots, \Omega_{\eqone}(x), \ldots) \to C[\funtwo(\ldots,
  \Omega'_{\eqone}(x), \ldots)]$$ and normalization implies that
  $\size{\Omega'_{\eqone}} \leq \size{\Omega_{\eqone}}$ (previous
  lemma).

  So, in our case, we have:
  \begin{eqnarray*}
  \valthree &=& \Omega_{\alpha}(\varone) \to \Omega'_{\alpha}(\varone) =
  \valone = \Omega_{\beta}(\varone') \to \Omega'_{\beta}(\varone')\\
  &=&\valone' = \Omega_{\gamma}(\varone'') \to \Omega'_{\gamma}(\varone'')
  = \valone''\\
  \valthree&=&\Omega_{\beta}(\vartwo) \to \Omega'_{\beta}(\vartwo) =
  \valtwo = \Omega_{\alpha}(\vartwo') \to \Omega'_{\alpha}(\vartwo')\\
  &=&\valtwo' = \Omega_{\gamma}(\vartwo'') \to \Omega'_{\gamma}(\vartwo'')
  = \valtwo''
  \end{eqnarray*}

  Because of normalization, $\size{\Omega_{\beta}} \geq
  \size{\Omega'_{\alpha}}$. Hence $\varone'$ is a suffix of $\varone$,
  itself a suffix of $\valthree$. Similarly, $\varone''$ and
  $\vartwo''$ are suffixes of $\valthree$.

  Since they're both suffixes of the same word, it is sufficient to show
  that they have the same length in order to show that they are
  identical. 
  \begin{eqnarray*}
    \size{\varone} & = & \size{\valthree} - \size{\Omega_{\alpha}}\\
    \size{\varone'} & = & \size{\valone} - \size{\Omega_{\beta}} =
    \size{\valthree} - \size{\Omega_{\alpha}} + \size{\Omega'_{\alpha}}
    - \size{\Omega_{\beta}}\\
    \size{\varone''} & = & \size{\valthree} - \size{\Omega_{\alpha}} +
    \size{\Omega'_{\alpha}} - \size{\Omega_{\beta}} +
    \size{\Omega'_{\beta}} - \size{\Omega_{\gamma}}\\
    \size{\vartwo''} & = & \size{\valthree} - \size{\Omega_{\beta}} +
    \size{\Omega'_{\beta}} - \size{\Omega_{\alpha}} +
    \size{\Omega'_{\alpha}} - \size{\Omega_{\gamma}}
  \end{eqnarray*}

  So $\varone'' = \vartwo''$ and thus $\valone'' = \valtwo''$.
\end{proof}

Proof of Corollary~\ref{cor:commute}:
\begin{proof}
  This is a generalization of the previous proof, not an induction on
  it. If $\omega_1 = \alpha_1 \ldots \alpha_n$ then the size in the last
  node is:
  $$ \size{\varone'} = \size{\valthree} - \sum \size{\Omega_{\alpha_i}}
  + \sum \size{\Omega'_{\alpha_i}} - \size{\Omega_{\alpha}}$$ which is
  only dependent on the commutative image of $\omega_1$.
\end{proof}

Proof of Proposition~\ref{prop:eppo-cd-bound}:
\begin{proof}
  Any descendant of $\nodeone$ with the same precedence can be
  identified with a word over $\{\funone^1, \ldots, \funone^M\}$. By
  Corollary~\ref{cor:branch} we know that these words have length at
  most $I$ and by Corollary~\ref{cor:commute} we know that it is
  sufficient to consider these words modulo commutativity.

  Modulo commutativity, words can be identified to vectors with as many
  components as the number of letters in the alphabet and whose sum of
  components is equal to the length of the word.

  Let $D^n_i$ be the number of elements from $\NN^n$ whose sum is
  $i$. This is the number of words of length $i$ over a $n$-ary alphabet
  modulo commutativity.

  Clearly, $D_{i-1}^n \leq D_i^n$ (take all the $n$-uple whose sum is
  $i-1$, add $1$ to the first component, you obtain $D_{i-1}^n$
  different $n$-uple whose sum of components is $i$).

  Now, to count $D_i^n$, proceed as follows: Choose a value $j$ for the
  first component, then you have to find $n-1$ components whose sum is
  $i-j$, there are $D_{i-j}^{n-1}$ such elements.

  \begin{eqnarray*}
    D_i^n &=& \sum_{0 \leq j \leq i} D_{i-j}^{n-1} = \sum_{0 \leq j \leq i}D_{j}^{n-1} \\
      &\leq& (i+1) \times D_i^{n-1} \leq (i+1)^{n-1} \times D_i^1\\
      &\leq& (i+1)^n
  \end{eqnarray*}
\end{proof}
}

\condinc{}{
\section{Semi-lattices of Quasi-Interpretations}\label{sect:semi-lattice}

The study of necessary conditions on programs satisfying the P-criterion
has drawn our attention to uniform quasi-interpretations. This suggests
to consider quasi-interpretations with fixed assignments for constructors
and to examine their properties as a class.

\begin{definition}[Compatible assignments]
  Let $\progone$ be a program and $\qi{.}_1$, $\qi{.}_2$ be two
  assignments for $\progone$. We say that they are \textit{compatible} if
  for any constructor symbol $\conone$ we have:
  $$\qi{\conone}_1=\qi{\conone}_2.$$.
  A family of assignments for $\progone$ is compatible if its elements
  are pairwise compatible.
  We use these same definitions for quasi-interpretations.
\end{definition}

Each choice of assignments for constructors thus defines a
\textit{maximal compatible family} of quasi-interpretations for a
program $\progone$: all quasi-interpretations for $\funone$ which take
these values on $\Constructors$.
 
We consider on assignments the extensional order $\leq$:
$$
\begin{array}{l}
\qi{.}_1 \leq \qi{.}_2 \quad \mbox{ iff }\\
 \quad \forall f \in
\Constructors \cup \Functions, \forall \vec{x} \in ({\RR}^{+})^k,
\qquad \qi{f}_1(\vec{x}) \leq \qi{f}_2(\vec{x}).
\end{array}$$

Given two compatible assignments $\qi{.}_1$, $\qi{.}_2$ we denote by
$\qi{.}_1 \wedge \qi{.}_2$ the assignment $\qi{.}_0$ defined by:
\begin{eqnarray*}
\forall c\in\Constructors.\qi{c}_0&=& \qi{c}_1= \qi{c}_2\\
\forall f\in\Functions.\qi{f}_0&=&\qi{f}_1\wedge\qi{f}_2
\end{eqnarray*} 
where $\alpha\wedge\beta$ denotes 
the greatest lower bound of $\{\alpha,\beta\}$ in
the pointwise order. Then we have:
\begin{proposition}\label{glbof2qi}
  Let $\funone$ be a program and $\qi{.}_1$, $\qi{.}_2$ be two
  quasi-interpretations for it, then $\qi{.}_1 \wedge \qi{.}_2$ is also
  a quasi-interpretation for $\funone$.
\end{proposition}

To establish this Proposition we need intermediary Lemmas. We continue
to denote $\qi{.}_0=\qi{.}_1 \wedge \qi{.}_2$:
\begin{lemma}\label{monotonicitylemma}
  For any $f$ of $\Functions$ we have that $\qi{f}_0$ is monotone and
  satisfies the subterm property.
\end{lemma}
\begin{proof}
  To prove monotonicity, assume $\vec{x}\leq \vec{y}$, for the product
  ordering.  Then, for $i=1$ or $2$: $ \qi{f}_0(\vec{x})=
  \qi{f}_1(\vec{x})\wedge \qi{f}_2(\vec{x}) \leq \qi{f}_i(\vec{x}) \leq
  \qi{f}_i(\vec{y}) $, using monotonicity of $\qi{f}_i$.
  As this is true for $i=1$ and $2$ we thus have: $
  \qi{f}_0(\vec{x})\leq \qi{f}_1(\vec{y})\wedge \qi{f}_2(\vec{y})=
  \qi{f}_0(\vec{y})$.
  It is also easy to check that $\qi{f}_0$ satisfies the subterm
  property.
\end{proof}

\begin{lemma}\label{boundQI0}
  Let $t$ be a term. We have: $\qi{t}_0\leq \qi{t}_i$, for $i=1, 2$.
\end{lemma}
\begin{proof}
  By induction on $t$, using the definition of $\qi{f}_0$ and
  $\qi{c}_0$, and the monotonicity property of Lemma
  \ref{monotonicitylemma}.
\end{proof}

\begin{lemma}\label{conditionrules}
  Let $g(\many{\patone}{n}) \to \termone$ be an equation of the program $f$.
  We have:
  $$ \qi{g}_0\circ (\qi{p_1}_0, \dots, \qi{p_n}_0) \geq \qi{t}_0.$$ 
\end{lemma}
\begin{proof}
  As patterns only contain constructor and variable symbols, and by
  definition of $\qi{.}_0$, if $p$ is a pattern we have: $\qi{p}_0=
  \qi{p}_1= \qi{p}_2$.
  Let $i=1$ or $2$; we have:
  \begin{eqnarray*}
    \qi{g}_i (\qi{p_1}_i(\vec{x}), \dots, \qi{p_n}_i(\vec{x})) &\geq &
    \qi{t}_i(\vec{x})\\
&& \quad \mbox{ because $\qi{.}_i$ is a
      QI,}\\
    &\geq & \qi{t}_0(\vec{x})\\
&& \quad \mbox{ with Lemma \ref{boundQI0}.} 
  \end{eqnarray*}
  So:
  $$\qi{g}_i (\qi{p_1}_0(\vec{x}), \dots, \qi{p_n}_0(\vec{x})) \geq 
  \qi{t}_0(\vec{x}), \mbox{ as } \qi{p_j}_i= \qi{p_j}_0. $$ Write
  $\vec{y}= (\qi{p_1}_0(\vec{x}), \dots, \qi{p_n}_0(\vec{x}))$.  As
  $\qi{g}_1 (\vec{y}) \geq \qi{t}_0(\vec{x})$ and $\qi{g}_2 (\vec{y})
  \geq \qi{t}_0(\vec{x})$, by definition of $\qi{g}_0$ we get $\qi{g}_0
  (\vec{y}) \geq \qi{t}_0(\vec{x})$, which ends the proof.
\end{proof}

Now we can proceed with the proof of Prop. \ref{glbof2qi}:

\begin{proof} [ Proof of Prop. \ref{glbof2qi}]
  Observe that Lemma \ref{monotonicitylemma} ensures that $\qi{.}_0$
  satisfies the monotonicity and the subterm conditions, and Lemma
  \ref{conditionrules} that it satisfies the condition w.r.t. the equations
  of the program. The conditions for the constructors are also satisfied
  by definition.  Therefore $\qi{.}_0$ is a quasi-interpretation.
\end{proof}

\begin{proposition}
  Let $\funone$ be a program and $\mathcal{Q}$ be a family of compatible
  quasi-interpretations for $\funone$, then $\wedge_{\qi{.} \in
    \mathcal{Q}} \qi{.}$ is a quasi-interpretation for $\funone$.
  Therefore maximal compatible families of quasi-interpretations for
  $\funone$ have an inferior semi-lattice structure for $\leq$.
\end{proposition}
\begin{proof}
  It is sufficient to generalize Lemmas \ref{monotonicitylemma} and
  \ref{conditionrules} to the case of an arbitrary family $\mathcal{Q}$
  and to apply the same argument as for the proof of Prop.
  \ref{glbof2qi}.
\end{proof}}

\end{document}